\documentclass[a4paper,11pt]{article}
\usepackage{jheppub}
\pdfoutput=1 
\usepackage{hyperref}
\usepackage{graphicx}
\usepackage{amsmath}
\usepackage{amsthm}
\usepackage{amssymb}
\usepackage{tipa}
\usepackage[caption=false]{subfig}
\usepackage{xcolor}
\usepackage{color}
\usepackage{mathrsfs}
\usepackage{tikz}
\usetikzlibrary{shapes.geometric}
\usepackage{tikz-cd}
\usetikzlibrary{arrows}
\usetikzlibrary{intersections}

\theoremstyle{plain}

\newcommand{\nn}{\nonumber}
\newcommand{\QTr}{\widetilde{\operatorname{Tr}}}
\newcommand{\Tr}{\operatorname{Tr}}
\renewcommand{\bar}{\overline}
\renewcommand{\tilde}{\widetilde}

\begin{document}

\title{Fractional Statistics and the Butterfly Effect}
\author{Yingfei Gu}
\author{and Xiao-Liang Qi}
\affiliation{Department of Physics, Stanford University, Stanford, CA 94305, USA}
\emailAdd{yfgu@stanford.edu}
\emailAdd{xlqi@stanford.edu}


\abstract{Fractional statistics and quantum chaos are both phenomena associated with the non-local storage of quantum information. 
In this article, we point out a connection between the butterfly effect in (1+1)-dimensional rational conformal field theories and fractional statistics in (2+1)-dimensional topologically ordered states. This connection comes from the characterization of the butterfly effect by the out-of-time-order-correlator proposed recently. We show that the late-time behavior of such correlators is determined by universal properties of the rational conformal field theory such as the modular S-matrix and conformal spins. Using the bulk-boundary correspondence between rational conformal field theories and (2+1)-dimensional topologically ordered states, we show that the late time behavior of out-of-time-order-correlators is intrinsically connected with fractional statistics in the topological order. We also propose a quantitative measure of chaos in a rational conformal field theory, which turns out to be determined by the topological entanglement entropy of the corresponding topological order.}

\maketitle

\section{Introduction}

In classical chaos theory, the butterfly effect refers to the exponential sensitivity of the state of the system to initial conditions, which can be measured quantitatively by various quantities, such as the Lyapunov exponents. For example, denote one of the coordinates of the chaotic system as $x(t)$, which is a function of the initial conditions. The butterfly effect occurs for small perturbation to the initial position $x(0)\rightarrow x(0)+\delta x(0)$ if the change $|\delta x(t)|\propto e^{\lambda_L t} |\delta x(0)| $ grows exponentially in time, with $\lambda_L$ the Lyapunov exponent. The variation $\frac{\delta x(t)}{\delta x(0)}$ can be calculated by the Poisson braket $\{x(t),p(0)\}_{\text{PB}}$. More generally, to obtain all Lyapunov exponents one should study the Poisson brakets $\{q_i(t),q_j(0)\}_{\text{PB}}$ with $q_i$ different components of coordinates and canonical momenta. 

The Poisson bracket formula suggests a natural generalization to quantum systems. In Heisenberg picture, quantum chaos can be characterized by the growth of operators of the form $i\left[x(t),p(0)\right]$. More precisely, for a given density matrix $\rho$ of a quantum particle, one should study $-{ \Tr}\left(\rho\left[x(t),p(0)\right]^2\right)$, which measures the size of the commutator in the state $\rho$.\footnote{This type of measure of chaos first appears in the semi-classical treatment of a superconductor by Larkin and Ovchinnikov\cite{larkin1969quasiclassical}}  (It should be noted that the square of the commutator should be considered since we are interested in the size of the operator.)
Recently, a generalization of such quantities have been studied in many-body systems, where operators $x$ and $p$ are replaced by generic many-body operators.\cite{shenker2014black,kitaev2014hidden,
maldacena2015bound}. If we take the thermal equilibrium state of the many-body system, the quantum butterfly effect refers to the increase of the thermal expectation value of the ``commutator norm square'' :
\begin{equation}
C(t):=\langle  |[W(t),V(0)]|^2  \rangle_\beta
\end{equation}
where $W(t)$ and $V(0)$ are generic Heisenberg operators at time $t$ and $0$. The thermal expectation value $\langle \text{-} \rangle_\beta$ is evaluated via the trace: $C(t)=Z^{-1} \Tr \left( e^{-\beta H} |[W(t), V(0)]|^2 \right)$. When expanding the ``commutator norm square,'' there are four terms in the function $C(t)$:
\begin{align}
C(t)= \langle V^\dagger(0) W^\dagger(t) &W(t) V(0) + V^\dagger(0) W^\dagger(t) W(t) V(0)\nn \\ &\underbrace{-  W^\dagger(t) V^\dagger(0) W(t) V(0) - V^\dagger(0) W^\dagger(t) V(0) W(t) }_{\text{``Out-of-time-ordered correlators''}}\rangle_\beta
\end{align}
The first two terms have the time order that appears in the response functions, therefore, are ``accessible''\cite{kitaev2014hidden}. However, the last two terms have a special ``out-of-time-order'' that is difficult to measure in conventional experiments. In a generic quantum many-body system, we expect the thermal averages of the accessible correlators to approach constant after the thermal time scale $\beta$; while the out-of-time-ordered correlators (OTOCs), 
possesses a non-trivial time dependence: it starts at large value and then decreases. The decrease of OTOCs corresponds to the increase of $C(t)$; therefore, it characterizes the quantum butterfly effect.

The OTOC has been extensively discussed in the connection between quantum gravity and quantum chaos, see Ref.~\cite{shenker2014black,kitaev2014hidden,
 shenker2014multiple, roberts2015localized, shenker2015stringy}. In those settings, for generic operators $W$ and $V$, the OTOC stays large until ``scrambling time''\cite{page1993average, hayden2007black, sekino2008fast} $t_{scr}$, then it decreases rapidly to $0$ and stays at $0$ in late time $t\gg t_{scr}$. Those behaviors are indicated by gravity, and are expected for a \emph{strongly} chaotic quantum system. On the other hand, also interesting is the measurement of OTOC in ``less chaotic'' or ``non-chaotic'' models (e.g. see Ref.~\cite{stanford2015many,michel2016four}) to see how different aspects of the model affect the chaos. In this paper, we measure OTOCs in $(1+1)$-dimensional rational conformal field theories (RCFTs)\footnote{For review, see Ref.~\cite{moore1990lectures}}, which are known to be integrable. We focus on the late time regime $t\gg t_{scr}$ and see how the ``integrability'' stops the late time value from vanishing in RCFTs. Remarkably, our result shows that when time $t$ goes to $\infty$, OTOCs in $(1+1)$-dimensional RCFTs are intrinsically related to the fraction statistics\cite{leinaas1977theory,wilczek1982quantum} in $(2+1)$-dimensional topological order. 
 
Fractional statistics and chaos are both interesting phenomena associated with non-local storage of information in strongly interacting quantum many-body systems. Fractional statistics usually arise in two-dimensional topological ordered states such as fractional quantum Hall states (FQH)\cite{tsui1982two,laughlin1983anomalous}. Theoretically, one can model the edge of FQH states using rational conformal field theories that possess the same algebraic structures/topological order as the corresponding FQH states. We use this type of bulk-boundary correspondence\footnote{This is the type of correspondence between $(1+1)$-dimensional RCFTs and $(2+1)$-dimensional TQFTs, e.g. WZW/CS correspondence. One should avoid confusing with the AdS/CFT correspondence.} to build the connections between the butterfly effect on the boundary and the fractional statistics in the bulk.

In section 2, we present the CFT computation of OTOC in the content of RCFTs, following the technique developed in Ref.~\cite{roberts2015diagnosing}, and show that the OTOC at $t=\infty$ only depends on the \emph{modular S-matrix}. Section 2 also includes more examples and discussion of small late time values. Section 3 is devoted to an alternative derivation of the result via bulk-boundary correspondence, which shows the non-trivial ``topology'' behind the OTOC. We also consider the OTOC between ``random operators'' in section 4, where we show an unexpected connection between the boundary butterfly effect and the bulk topological entanglement entropy. Section 5 contains conclusions and discussions on real condensed matter experiments. Most of the technical details are placed in the appendices.

\section{Out-of-time-ordered-correlators in rational conformal field theories}

\subsection{Definitions and conventions}

\begin{figure}[h]
\center
\subfloat[]{
\begin{tikzpicture}[scale=1.2,baseline={(current bounding box.center)},line width=0.8pt]
\filldraw[fill=gray] (0pt,0pt) ellipse (20pt and 3pt);
\draw (6pt,5pt)--(6pt,40pt);
\draw[>=stealth,->] (-6pt,40pt)--(-6pt,5pt);
\draw (6pt,40pt)..controls (6pt,60pt) and (-6pt,60pt)..(-6pt,40pt);
\node at (-11pt,20pt){$t$};
\node at (13pt,10pt){$V$};
\filldraw (6pt,10pt) circle (0.6pt);
\filldraw (6pt,40pt) circle (0.6pt);
\node at (18pt,40pt){$W(t)$};
\node at (0pt,-12pt){$|x\rangle =W(t)V| \beta \rangle $};
\end{tikzpicture}}\hspace{30pt}
\subfloat[]{
\begin{tikzpicture}[scale=1.2,baseline={(current bounding box.center)},line width=0.8pt]
\filldraw[fill=gray] (0pt,0pt) ellipse (20pt and 3pt);
\draw (6pt,5pt)--(6pt,40pt);
\draw[>=stealth,->] (-6pt,40pt)--(-6pt,5pt);
\draw (6pt,40pt)..controls (6pt,60pt) and (-6pt,60pt)..(-6pt,40pt);
\node at (-11pt,20pt){$t$};
\node at (0pt,10pt){$V$};
\filldraw (-6pt,10pt) circle (0.6pt);
\filldraw (6pt,40pt) circle (0.6pt);
\node at (18pt,40pt){$W(t)$};
\node at (0pt,-12pt){$|y \rangle = V W(t)| \beta \rangle $};
\end{tikzpicture}}
\caption{Illustration of the thermal expectation value of OTOC as an inner product of states $|x\rangle$ and $|y\rangle$. The operator $V$ and $W$ act on a pure state $|\beta \rangle$ (the grey disk) which is a purification of the thermal state. We can imagine $W(t)$ as a small perturbation in late time $t\gg \beta$: if $W={\rm Identity}$ is trivial, then state $|x\rangle=|y\rangle$; if $W$ is a non-trivial perturbation and the system is ``chaotic'', we expect the ``butterfly'' $W$ causes a big difference on states $|x\rangle$ and $|y\rangle$. Therefore, we can use the inner product $\langle y | x \rangle$ to quantify the butterfly effect.}
\label{fig:OTO as inner product}
\end{figure}
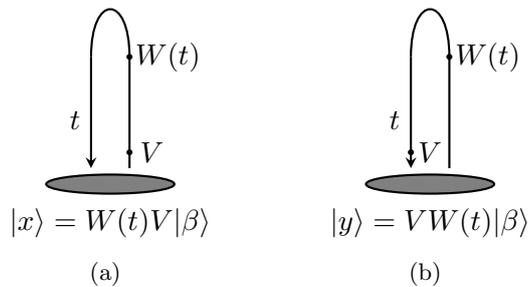

Before going to the detailed discussions, we first declare some general definitions for OTOC. More specifically, we are considering thermal expectation value of OTOC in a quantum system with Hilbert space $\mathcal{H}$:
$f(t):=\langle W^\dagger (t)V^\dagger (0)W(t)V(0) \rangle_\beta
$, $V$ and $W$ act on $\mathcal{H}$. Alternatively, one can interpret $f(t)$ as an inner product between two pure states (see figure~\ref{fig:OTO as inner product}):
\begin{eqnarray}
f(t)= \langle y | x\rangle; \quad |x\rangle = W(t)V|\beta \rangle,\ |y\rangle = VW(t)|\beta \rangle
\end{eqnarray} 
where $|\beta \rangle$ is a purification of the thermal system at temperature $T=1/\beta$, e.g., a thermofield double state\cite{israel1976thermo,maldacena2003eternal} $|\beta\rangle =Z^{-1/2} \sum_{n} e^{-\beta E_n/2} |n\rangle |\bar{n}\rangle \in \mathcal{H} \otimes \overline{\mathcal{H}}$. We require all operators act on one side, say $\mathcal{H}$.

As a measure of the difference between $|x\rangle$ and $|y\rangle$, it is convenient to use the normalized value:
\begin{eqnarray}
\tilde{f}(t):= \frac{\langle y | x \rangle }{\sqrt{\langle x | x\rangle \langle y | y \rangle}}
\end{eqnarray}

Previous studies\cite{shenker2014black,kitaev2014hidden, shenker2015stringy,roberts2015diagnosing} of such correlation function focused on the early time $\beta < t < t_{scr}$, i.e., between the dissipation time and the scrambling time, and found interesting ``Lyapunov behavior''\footnote{here Lyapunov behavior refers to the functional dependence of $f(t)$ on $t$ at early time :\[f(t) \sim 1- \frac{1}{N} e^{\lambda_L t} + \ldots  \] where $N$ is some big number in the model, and the exponent $\lambda_L$ is recognized as Lyapunov exponent. Physically, the Lyapunov behavior characterizes how fast the chaos develops in the system.} for those systems that can be holographically described by Einstein gravity. However, for a generic system that does not have a large separation between dissipation time $t_d \sim \beta$ and scrambling time $t_{scr}$, the ``Lyapunov behavior'' is not well-defined. Instead, one can focus on the later time regime $t\gg t_{scr} $, which characterized the residue part of the system that ``survived'' under the butterfly effect.

For later convenience, we denote the inner product $\langle	x | x \rangle = \langle V^\dagger W^\dagger (t) W(t) V \rangle_\beta$ by $g(t)$. In the regime $t\gg \beta$, such a four point function generically factorizes to $g(t)\sim \langle V^\dagger V \rangle_\beta \langle W^\dagger W \rangle_\beta$, which represents a normalization for operators $W$ and $V$. The same applies to $\langle y | y \rangle$. In the following section, we will study the late time behavior of $\tilde{f}(t)$ in the context of RCFTs.

\subsection{An overview of existing results}

As a starting point of our discussion, we briefly review the calculation of OTOC in $(1+1)$-d CFT in 
Ref.~\cite{roberts2015diagnosing}.

To setup the CFT computation, it is essential to use complexified time $t_c=t-i\tau$, where $t$ stands for the Minkowski time, and $\tau$ for the Euclidean time. The strategy is to begin with a pure Euclidean computation with $t=0$ and then {\it analytically continue} it to the desired Minkowski time $t$. Such strategy enables us to manipulate the order of operators by tuning the auxiliary small imaginary part $\tau=\epsilon$. After the conformal mapping $z=\exp\left({2\pi w}/{\beta} \right), \quad \bar{z}=\exp \left( {2\pi \bar{w}}/{\beta}\right)$, with $w=x-t_c$ and $\bar{w}=x+t_c$\footnote{We use a different sign convention comparing to Ref.~\cite{roberts2015diagnosing} for later convenience.}, the thermal expectation value is mapped to the vacuum expectation value. In the end, it is essential to consider the vacuum expectation value of the four point function which has a general decomposition in terms of conformal blocks\cite{belavin1984infinite,francesco2012conformal,roberts2015diagnosing} 
\begin{eqnarray}
\langle W^\dagger (z_1,\bar{z}_1) W(z_2,\bar{z}_2) V^\dagger(z_3,\bar{z}_3) V(z_4,\bar{z}_4) \rangle =\frac{1}{z_{12}^{2h_w} z_{34}^{2h_v} } \frac{1}{\bar{z}_{12}^{2\bar{h}_w} \bar{z}_{34}^{2\bar{h}_v} } \sum\limits_{p,\bar{p}} g_{p,\bar{p}} \mathcal{F}_p\left(\eta\right) \overline{\mathcal{F}}_{\bar{p}}\left(\bar{\eta} \right)
\end{eqnarray}
where $z_{ij}=z_i-z_j$, and $\eta= \frac{z_{12}z_{34}}{z_{13}z_{24}},~\bar{\eta}= \frac{\bar{z}_{12} \bar{z}_{34}}{ \bar{z}_{13} \bar{z}_{24}}$ are cross ratios, $g_{p,\bar{p}}$ is the pairing coefficient for holomorphic block $p$ and anti-holomorphic block $\bar{p}$. It is important to note that the summation is {\bf only} over conformal families labeled by $\lbrace p,\bar{p} \rbrace$ rather than all the primaries and descendants individually. For the parameter regime $t\gg \beta$ that we are interested in, the cross ratio is always small: $\eta ,\bar{\eta} \sim \exp(-{2\pi t}/{\beta}) \ll 1$. Therefore, we can formally expand $\mathcal{F}_p(\eta)$ according to the powers of $\eta$: 
\begin{eqnarray}
\mathcal{F}_p(\eta) =\sum_{n=0}^\infty F_{p,n} \eta^{h_p+n}, \quad \textit{if} \ N_{a\bar{a}}^p N_{b\bar{b}}^p \neq 0
\label{eqn:conformal block power expansion}
\end{eqnarray}
where $h_p$ denotes the dimension of the primary $p$ (and $h_p+n$ for descendants, $n\geq1$). $\lbrace F_{p,n} \rbrace$ are coefficients depending on the details of operator $W$ and $V$. $a,b$ label the conformal families of $W,V$, correspondingly. (For simplicity we can assume $W,V$ are primaries. We sometimes make the family label explicit by denoting $W=W[a]$ and $V=V[b]$.) The fusion multiplicity $N_{a\bar{a}}^p$ counts the dimension of operator product algebra from $a,\bar{a}$ to intermediate channel $p$. $\mathcal{F}_p=0$ if $N_{a\bar{a}}^p N_{b\bar{b}}^p=0$, i.e., if $p$ is absent in the fusion channels of either $a\times \bar{a}$ or $b\times \bar{b}$.

The behavior of OTOC is determined by the dependence of conformal blocks on cross ratio $\eta$: for arbitrary four points on complex plane, it is convenient to use a conformal map to move three of them to standard positions: $0,1$ and $+\infty$, and leave over with a free parameter $\eta$. Therefore, the conformal blocks $\lbrace \mathcal{F}_p(\eta) \rbrace $ will have three singularities in general: $\eta=0,1$ and $\infty$ (see figure~\ref{subfig:cross ratio winding}). The same is true for $\overline{\mathcal{F}}(\bar{\eta})$. When implementing the analytic continuation to Minkowski time $t$, the path $\eta(t)$ might go around a singular point and leave from the principal sheet to the \emph{second sheet}. One can check (e.g. see Ref.~\cite{roberts2015diagnosing}) that among conformal blocks in the OTOC $f(t)$, only the holomorphic $\eta$ winds a topologically nontrivial loop\footnote{Technically speaking, to make the statement more precise, one need to place $W^\dagger$ and $W$ at $x\gg \beta$, s.t. $\eta$ starts at vicinity of $0$. This specific setting is not essential to the results in this paper. The key point is that the difference between path $\eta(t)$ and $\bar{\eta}(t)$ is a nontrivial winding around singularity at $z=1$, which is always valid for $f(t)$ at late time regardless of the initial position of operators.}, as is shown in figure~\ref{subfig:cross ratio winding}. In contrast, the winding of $\eta$ or $\bar{\eta}$ is trivial in the in-time-ordered correlator $g(t)$. This is the key difference that leads to a nontrivial ratio $\tilde{f}(t)$.

\subsection{Rational conformal field theories}

For general CFTs, the number of conformal families can be infinite and there is not much general information we can tell. One can instead restrict to certain subclass of CFTs. For example in Ref.~\cite{roberts2015diagnosing}, the authors analyzed the behavior of the Virasoro identity block of a holographic CFT\cite{hartman2013entanglement,fitzpatrick2014universality,asplund2015holographic} and deduced a result about the butterfly effect that is consistent with the holographic analysis\footnote{To avoid confusion, we clarify that the previous studies\cite{shenker2014black,roberts2015diagnosing,kitaev2014hidden} were mainly interested in the Lyapunov exponent and scrambling time, which are in the ``early time'' regime, while the current work will focus on the ``late time'' regime instead.}.

In this paper, we will choose to work in another rich subclass of CFTs---diagonal rational conformal field theories, which have a well-controlled algebraic structure on conformal blocks (for a review of RCFTs, see Ref.~\cite{moore1990lectures} and reference therein). In this subclass, the four point functions {in Euclidean time} have a simpler finite sum presentation: 
\begin{eqnarray}
\langle W^\dagger (z_1,\bar{z}_1) W(z_2,\bar{z}_2) V^\dagger(z_3,\bar{z}_3) V(z_4,\bar{z}_4) \rangle  =\frac{1}{z_{12}^{2h_w} z_{34}^{2h_v} } \frac{1}{\bar{z}_{12}^{2\bar{h}_w} \bar{z}_{34}^{2\bar{h}_v} } \sum\limits_{i=1}^N  \mathcal{F}_i\left(\eta\right) \overline{\mathcal{F}}_{i}\left(\bar{\eta} \right)
\end{eqnarray}
where holomorphic conformal blocks $\lbrace \mathcal{F}_i \rbrace$ form a {\bf finite} dimensional vector space: $V_{\bar{a} a \bar{b} b}$.\footnote{Here we use an index $i$ instead of the intermediate channel label $p$ due to the subtlety that a fixed channel $p$ might have dimension $N_{a\bar{a}}^p N_{b\bar{b}}^p>1$. Later we will ignore this subtlety and use indices $i,j$ to label the intermediate channels as well. Alternatively, one can treat channel labels as labeling a subspace instead of a vector.} This linear space is parametrized by the cross ratio $\eta$, and the same applies to the anti-holomorphic $\overline{\mathcal{F}}_i$ and $\bar{\eta}$.\footnote{Alternatively, one can formulate conformal blocks as vector bundle over the moduli space\cite{friedan1987analytic}. The monodromy is defined as that of the vector bundle. Such a vector bundle is also equipped with a fiber-wise metric for computing physical correlation functions. In the diagonal theory here, the metric is $\delta_{ij}$.}

\begin{figure}[h]
\center
\subfloat[Monodromy (analytic)]{
\begin{tikzpicture}[scale=1.2,baseline={(current bounding box.center)}]
\draw [>=stealth,->](0pt,-18pt)--(0pt,40pt);
\draw [>=stealth,->](-10pt,0pt)--(60pt,0pt);
\node at (60pt,35pt){$z$};
\draw (55pt,30pt)--(65pt,30pt);
\draw (55pt,30pt)--(55pt,40pt);
\filldraw (0pt,0pt) circle (0.8pt);
\node at (-2pt,-5pt){ $0$};
\filldraw (3pt,0pt) circle (0.8pt);
\node at (16pt,14pt){ $\eta(t)$};
\filldraw (22pt,0pt) circle (0.8pt);
\node at (22pt,-5pt){ $1$};
\filldraw (50pt,0pt) circle (0.8pt);
\node at (50pt,-5pt){ $\infty$};
\draw[densely dashed] (3pt, 0pt)..controls (12pt,12pt) and (25pt, 12pt)..(25pt,0pt);
\draw[>=stealth, ->, densely dashed] (25pt, 0pt)..controls (25pt,-12pt) and (12pt, -12pt)..(3pt,0pt);
\end{tikzpicture}
\label{subfig:cross ratio winding}
}
\hspace{20pt}
\subfloat[Monodromy (algebraic)]{
\begin{tikzpicture}[scale=1.2,baseline={(current bounding box.center)}]
\draw (0pt,0pt)--(15pt,25pt);
\draw (-15pt,25pt)--(0pt,-0pt);
\draw (-7.5pt,12.5pt)--(0pt,25pt);
\draw (0pt,0pt)--(0pt,-20pt);
\node at (0pt,-25pt){$b$};
\node at (-15pt, 30pt){$a$};
\node at (0pt,30pt){$\bar{a}$};
\node at (15pt, 30pt){$b$};
\node at (-10pt,5pt){$j$};
\node at (-25pt,5pt){$\widetilde{M}$};
\end{tikzpicture}
$:=$
\begin{tikzpicture}[scale=1.2,baseline={(current bounding box.center)}]
\draw (-7.5pt,12.5pt)..controls (-3.75pt,18.75pt) and (3.75pt,6.25pt)..(7.5pt,12.5pt);
\draw[line width=4pt,draw=white] (0pt,0pt)..controls (3.75pt,6.25pt) and (-1.5pt,12.5pt)..(0.75pt,15.75pt);
\draw (0pt,0pt)..controls (3.75pt,6.25pt) and (-1.5pt,12.5pt)..(0.75pt,15.75pt);
\draw (0.75pt,15.75pt)..controls (3pt, 20pt) and (7.5pt,12.5pt)..(15pt,25pt);
\draw [line width=4pt,draw=white] (7.5pt,12.5pt)..controls (9.25pt,19.75pt) and (0.75pt,18.75pt)..(0pt,25pt);
\draw (7.5pt,12.5pt)..controls (9.25pt,19.75pt) and (0.75pt,18.75pt)..(0pt,25pt);
\draw (-15pt,25pt)--(0pt,-0pt);
\draw (0pt,0pt)--(0pt,-20pt);
\node at (0pt,-25pt){$b$};
\node at (-15pt, 30pt){$a$};
\node at (0pt,30pt){$\bar{a}$};
\node at (15pt, 30pt){$b$};
\node at (-10pt,5pt){$j$};
\end{tikzpicture}$=\sum\limits_{i=1}^N \widetilde{M}_{ij}$
\begin{tikzpicture}[scale=1.2,baseline={(current bounding box.center)}]
\draw (0pt,0pt)--(15pt,25pt);
\draw (-15pt,25pt)--(0pt,-0pt);
\draw (-7.5pt,12.5pt)--(0pt,25pt);
\draw (0pt,0pt)--(0pt,-20pt);
\node at (0pt,-25pt){$b$};
\node at (-15pt, 30pt){$a$};
\node at (0pt,30pt){$\bar{a}$};
\node at (15pt, 30pt){$b$};
\node at (-10pt,5pt){$i$};
\end{tikzpicture}}
\caption{(a) Winding $\eta(t)$ in the moduli space: $\eta(t)\in \mathbb{C}-\lbrace0,1,+\infty \rbrace$ winds around $z=1$ clockwisely and induces a linear transformation on the space of conformal blocks $V_{\bar{a}a\bar{b}b}\simeq V_{b}^{a\bar{a}b}$, which defines the the monodromy matrix. (b) The diagrammatic representation of the monodromy matrix $\widetilde{M}= \widetilde{M}[a,b]$. For fixed $a,b$, $\widetilde{M}[a,b]$ acts by braiding the two lines $\bar{a},b$. 
See appendix~\ref{appendix: notations and conventions} for detailed conventions of the diagrammatics.}
\label{fig:analytic monodromy to diagrams}
\end{figure}
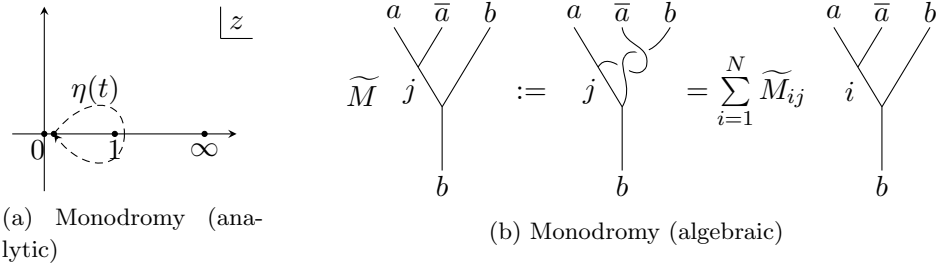

In this setting, we are able to discuss the effect of analytic continuation to the second sheet more concretely: according to the general principle of RCFTs\cite{moore1989classical}, a full winding around a singularity induces a linear transformation $\tilde{M}[a,b]$ in space $V_{\bar{a} a\bar{b} b}$, known as the {\it monodromy}\cite{belavin1984infinite}. See figure~\ref{fig:analytic monodromy to diagrams} for its definition in terms of diagrams.

With this definition, one can compute the matrix element of $\widetilde{M}$ in terms of the general algebraic data of RCFT, known as the F-matrix and the R-matrix. This computation is well-known in the literature, and we include a summary of it in the appendix~\ref{appendix: monodromy}. The monodromy matrix is determined by {\it universal data} that only depends on the algebraic structure of RCFTs, independent of the particular state chosen in the family and the details of locations of the operators. Now with $\widetilde{M}_{ij}$ defined in figure~\ref{fig:analytic monodromy to diagrams}, we can rewrite the formula of normalized OTOC more explicitly:
\begin{eqnarray}
\tilde{f}(t)=\frac{\sum\limits_{i,j=1}^N \widetilde{M}_{ij} \mathcal{F}_i(\eta) \overline{\mathcal{F}}_j(\bar{\eta})}{\sum\limits_{i=1}^N \mathcal{F}_i(\eta) \overline{\mathcal{F}}_i(\bar{\eta})}
\label{eqn:Rcft f(t)}
\end{eqnarray}
the normalization cancels the unimportant common prefactor $\frac{1}{z_{12}^{2h_w} z_{34}^{2h_v} }\frac{1}{\bar{z}_{12}^{2\bar{h}_w} \bar{z}_{34}^{2\bar{h}_v} }$. Nevertheless, the actual functional form of $\tilde{f}(t)$ is still messy because of the appearance of descendants when one explicitly expand the conformal blocks as power series of $\eta$. However, the result is simplified in the late time $t\gg \beta$ regime, with $\eta,\bar{\eta} \sim \exp(-2\pi t/\beta)$.

First of all, the residue value at $t\rightarrow \infty$ is free of complicated coefficients (i.e., those $F_{i,n}$s): 
\begin{eqnarray}
r[a,b]=\lim_{t\rightarrow \infty} \tilde{f}(t)= \frac{ \widetilde{M}[a,b]_{11} |\mathcal{F}_1(0)|^2 }{ |\mathcal{F}_1(0)|^2}=
\widetilde{M}[a,b]_{11}\label{eq:monodromy}
\end{eqnarray}
only depends on universal data: $(1,1)$ component of monodromy matrix $\widetilde{M}[a,b]$. Interestingly, the $(1,1)$ element of monodromy matrix can be expressed solely by the modular S-matrix (see appendix~\ref{appendix: monodromy} for a more detailed derivation):
\begin{eqnarray}
\widetilde{M}[a,b]_{11}=\frac{\mathcal{D} s_{ab}^*}{d_ad_b}=\frac{s_{11} s_{ab}^*}{s_{1a}s_{1b}}
\label{eqn: residue value of OTOC}
\end{eqnarray}
where $\mathcal{D}=1/s_{11}$ is the total quantum dimension, and $d_a=s_{1a}/s_{11}$ are the quantum dimensions of individual conformal families. (As a side remark, $\widetilde{M}[a,b]_{11}$ is known as {\it monodromy scalar} in the literature of anyon interferometry\cite{bonderson2007decoherence,overbosch2001inequivalent}, which we will discuss more in later part of the article.)

From the equation~(\ref{eqn: residue value of OTOC}), one can immediately see that Abelian theories\footnote{we borrow the terminology from anyon theory\cite{kitaev2006anyons,nayak2008non} that non-Abelian anyons $a,b$ refer to those conformal families having multiple fusion channels e.g. $a\times b= c + d + \ldots$, and Abelian theories refer to a theory without non-Abelian anyons.} have $|r|=1$. In general, one can prove a physically intuitive inequality $|r|\leq 1$ by noticing
\begin{eqnarray}
|s_{ab}|=\frac{1}{\mathcal{D}} \left\rvert \sum_c d_c  N_{ab}^c\frac{\theta_c}{\theta_a\theta_b} \right\rvert \leq \frac{1}{\mathcal{D}}  \sum_c d_c N_{ab}^c = \frac{d_a d_b}{\mathcal{D}}
\end{eqnarray}
 Therefore, $|r|= \mathcal{D} |s_{ab}|/d_a d_b \leq 1$. When there is only one fusion channel in $a\times b$, i.e. when $N_{ab}^c$ is nonzero for only one $c$, the inequality picks equal sign. 

From this somewhat ``trivial'' inequality, we see an interesting indication: ``scrambling''\footnote{Scrambling is defined as the process where a non-entangled state evolves into an almost maximally entangled state, or correspondingly, the process of a simple local operator evolves into a complicated operator that is supported on almost the whole system. For details see Ref.~\cite{shenker2014black,page1993average, hayden2007black, sekino2008fast}}, which refers to the suppression of $r$ here, only comes from non-Abelian anyons. Physically, the scrambling occurs due to nontrivial interference between different fusion channels, which describes the spreading of quantum information more and more non-locally among different conformal families. For Abelian anyons, the long-time value of OTOC is only different from the short-time value by a phase. Although such a phase is still a topological feature, it does not lead to scrambling since there is no nontrivial unitary transformation in the vector space generated. This observation is closely related to the correspondence between scrambling and fractional statistics, which we will elaborate in next section. 


Now we return to more general components of matrix $\widetilde{M}$, which appears in the form of $\widetilde{M}_{ij}\eta^{h_i+n} \bar{\eta}^{h_j+m}$ in the numerator of equation~(\ref{eqn:Rcft f(t)}). $n,m\in \mathbb{Z}^{\geq 0}$ come from possible descendants. As a function of $t$, such terms decay exponentially as $\widetilde{M}_{ij} \exp(-2\pi ({h_i+h_j+m+n})t/\beta)$. Except the $(1,1)$ element we mentioned before, it is generally unpractical to separate a prefactor from an exponential decaying function. However, we can still make non-trivial prediction on the ``spectroscopy'' of function $\tilde{f}(t)$ by knowing whether certain $\widetilde{M}_{ij}$ is non-zero. The exponents from the denominator: $\sum_{i=1}^N \mathcal{F}_i(\eta) \overline{\mathcal{F}}_i(\bar{\eta})$ is always in form of $\Delta_{i,n}=2h_i+n, n\in \mathbb{Z}^{\geq 0}$ as a consequence of diagonal pairing. However, the numerator: $\sum_{i,j=1}^N \widetilde{M}_{ij} \mathcal{F}_i(\eta) \overline{\mathcal{F}}_j(\bar{\eta})$ contains mixed exponents $\Delta_{i,j,n}=h_i+h_j+n$, which enriches the spectrum. In particular, the slowest decaying rate comes from pairing between identity block with the smallest scaling dimension $h_z$ (assuming $h_z<1$), if the corresponding matrix element $\widetilde{M}_{1z}$ (or $\widetilde{M}_{z1}$) is non-vanishing. Such elements of the monodromy matrix are related to the  ``generalized'' modular S-matrix $s_{\bar{z},a{\bar{b}}}$ for one-punctured-torus\cite{moore1988polynomial,kitaev2006anyons} as $\widetilde{M}_{1z}=\frac{\mathcal{D} s_{z,a\bar{b}}}{d_a d_b}$. More details are presented in appendix \ref{appendix: monodromy}.

\subsection{Examples}\label{subsec:example}

To make the abstract discussion more transparent, in this section we present three examples: (1) Ising model, see table~\ref{table: ising}; (2) Compactified boson, see table~\ref{table: compactified boson}; and (3) $\operatorname{SU}(2)$ WZW model at level $k$, see table~\ref{table: WZW models}.

\begin{table}[h]
\center
\begin{tabular}{|l|}
\hline
{\bf Ising model}\\
Primaries: $1, \sigma, \epsilon$;\\
Fusion rules: $\sigma \times \sigma = 1 + \epsilon$, $\epsilon \times \sigma = \sigma $, $\epsilon \times \epsilon = 1$;\\
S-matrix (with columns $(1,\sigma,\epsilon)$) and residue value: \\
$S=(s_{ab})=\frac{1}{2}
\begin{pmatrix}
1 & \sqrt{2} & 1 \\
\sqrt{2} & 0 & -\sqrt{2} \\
1 & -\sqrt{2} & 1 \\
\end{pmatrix}$,\\
$r[\sigma,\sigma]=0,\quad r[\sigma,\epsilon]= -1,\quad r[\epsilon,\epsilon]=1$.\\
\hline
\end{tabular}
\caption{Algebraic data of Ising model. Residue value $r$ is computed according to equation~(\ref{eq:monodromy}) and (\ref{eqn: residue value of OTOC}).
}
\label{table: ising}
\end{table}

The Ising model was considered as an example of non-chaotic theory in Ref.~\cite{roberts2015diagnosing}. Our result on late time residue value is consistent with Ref.~\cite{roberts2015diagnosing}. Method here has the advantage that it only relies on a small amount of universal data rather than the explicit functional form of conformal blocks or four point functions.

\begin{table}[h!]
\center
\begin{tabular}{|l|}
\hline
{\bf Compacified boson}\\
Primaries: $V_0,V_1,\ldots,V_{2N-1}$, $N\in \mathbb{Z}$;\\
Fusion rules: $V_i \times V_j = V_{i+j} $ ($i,j\in \mathbb{Z}$ mod $2N$);\\
S-matrix and residue value ($ 0\leq j,k \leq 2N-1$):\\
$s_{jk}=\frac{1}{\sqrt{2N}} \exp \left( \frac{jk }{2N} 2\pi i \right)$,\\
$r[V_j,V_k]= \exp \left( \frac{jk }{2N} 2\pi i \right)$.\\
\hline
\end{tabular}
\caption{Algebraic data of compactified boson. Residue value $r$ is computed according to equation~(\ref{eq:monodromy}) and (\ref{eqn: residue value of OTOC}).
}
\label{table: compactified boson}
\end{table}

The compactified boson is an example of Abelian theory. The residue value for this model is a pure phase $|r|=1$ since there is no non-Abelian anyon.

\begin{table}[h!]
\center
\begin{tabular}{|l|}
\hline
{\bf $\operatorname{SU}(2)$ WZW model at level $k$}\\
Primaries: $V_0,V_1,\ldots V_k$;\\
Fusion rules: $V_i \times V_j = \sum\limits_{l=|i-j|,~|i-j|+2,...}^{min(i+j,2k-(i+j))} V_l$;\\
S-matrix and residue value ($\ 0\leq i,j \leq k$):\\
$
s_{ij}=\sqrt{\frac{2}{k+2}} \sin \left(\frac{(i+1)(j+1)\pi}{k+2 }\right),
$\\
$
r[V_i,V_j]=\frac{\sin \left(\frac{\pi}{k+2 }\right)\sin \left(\frac{(i+1)(j+1)\pi}{k+2 }\right)}{\sin \left(\frac{(i+1)\pi}{k+2 }\right)\sin \left(\frac{(j+1)\pi}{k+2 }\right)}
$.\\
\hline
\end{tabular}
\caption{Algebraic data of $\operatorname{SU}(2)$ WZW model\cite{witten1984non,knizhnik1984current} at level k. Residue value $r$ is computed according to equation~(\ref{eq:monodromy}) and (\ref{eqn: residue value of OTOC}).}
\label{table: WZW models}
\end{table}

\begin{figure}[h!]
\center\subfloat{
\includegraphics[scale=0.5]{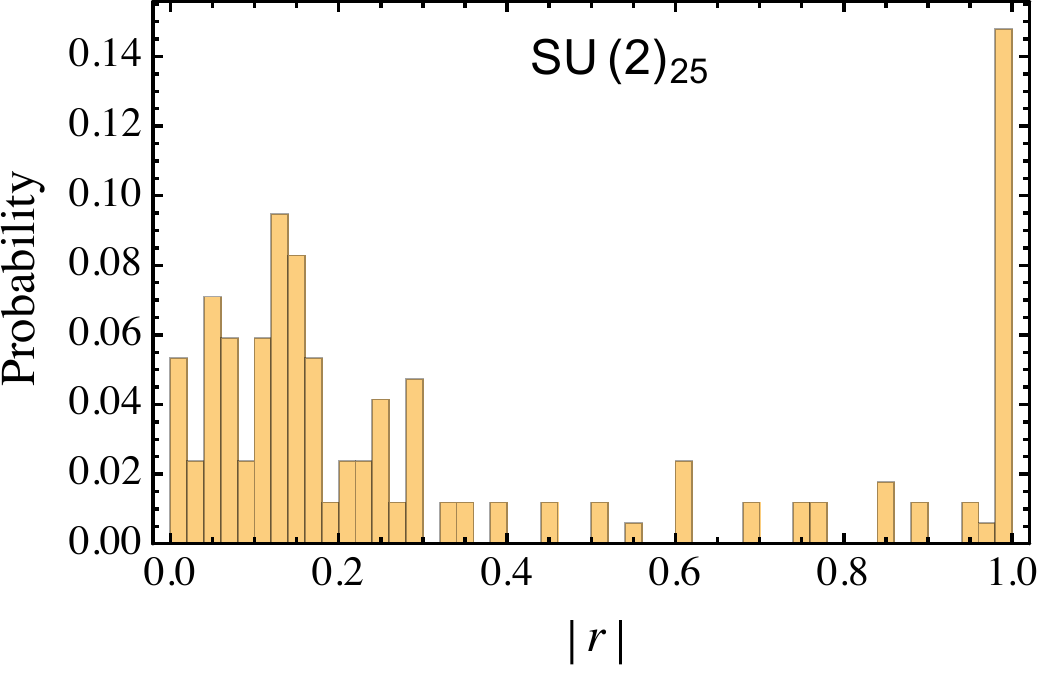}
}\hspace{15pt}
\subfloat{
\includegraphics[scale=0.5]{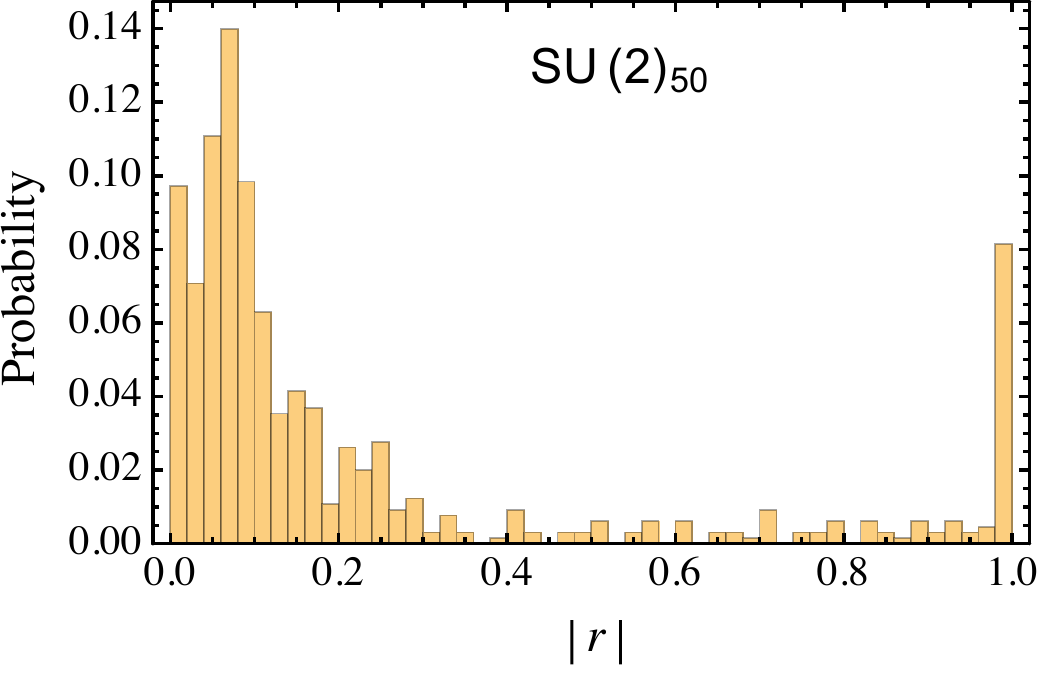}}

\subfloat{
\includegraphics[scale=0.5]{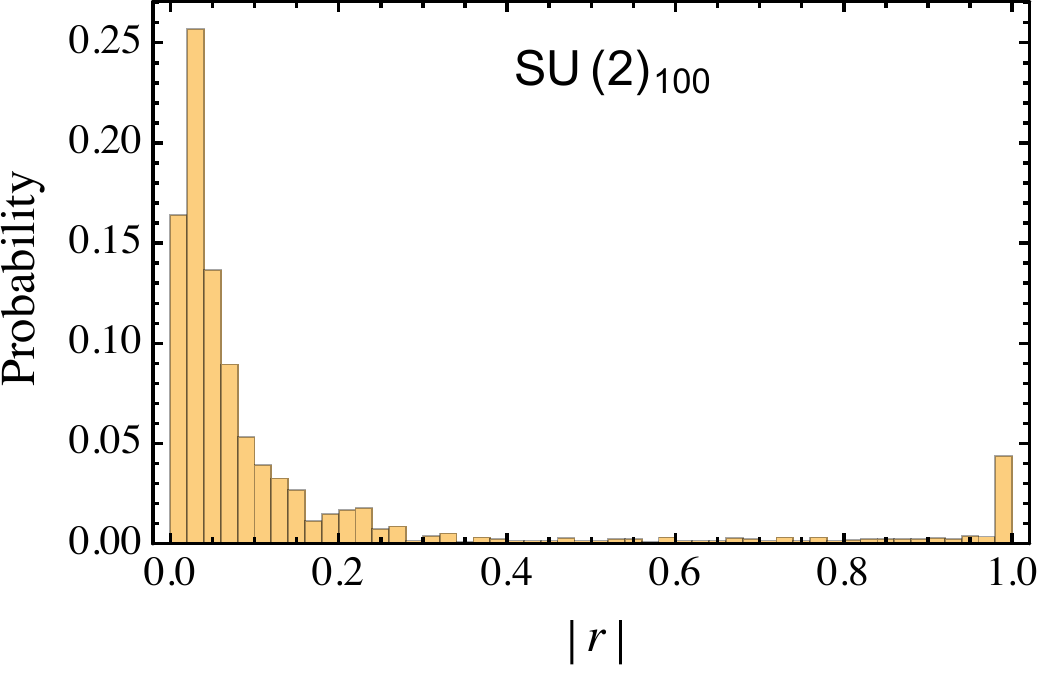}
}\hspace{15pt}
\subfloat{
\includegraphics[scale=0.5]{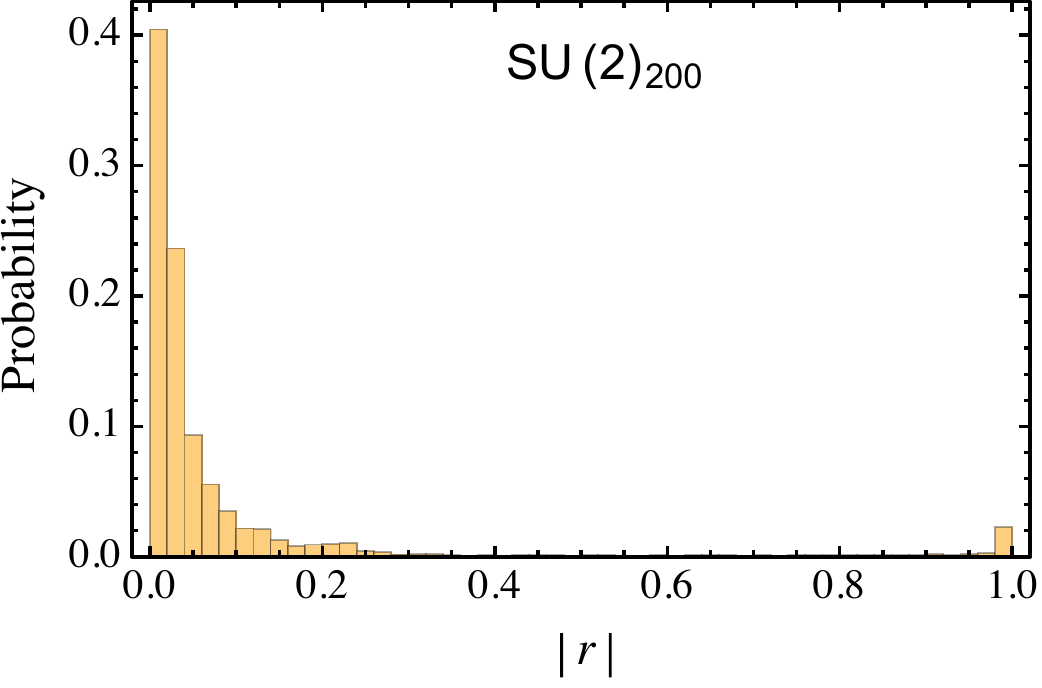}}
\caption{The distributions of $|r\left[V_i,V_j\right]|$ in $\operatorname{SU}(2)$ WZW models.}
\label{fig:su2levelk}
\end{figure}

In the example of WZW models, residue value $r$ is strongly oscillating in $i$ and $j$, so that some channels are more scrambled than others. To understand the overall behavior of the magnitude of $r$, which describes the strength of ``scrambling'', we show a probability distribution plot of $\left|r\left[V_i,V_j\right]\right|$ for the $\operatorname{SU}(2)_k$ WZW model at level $k=25,50,100,200$ in figure~\ref{fig:su2levelk}. One can see from the distribution that the theory is more chaotic in larger $k$, where most of the $|r|$ is close to zero.

In appendix~\ref{appendix: detailed WZW}, we present the statistics of $|r|$ for a related family of RCFT's, the ${\rm SU}(k)$ WZW models at level $2$. These theories are related to ${\rm SU}(2)_k$ by the level-rank duality\cite{francesco2012conformal}, but they have a large central charge $c=\frac{2(k^2-1)}{k+2}$. We prove there in appendix~\ref{appendix: detailed WZW} that the statistics of $|r|$ in ${\rm SU}(k)_2$ is identical to that of ${\rm SU}(2)_k$, which makes them interesting examples of RCFT's with stronger chaos in the large central charge limit. 

\section{The Bulk-boundary correspondence}

As can be seen from the discussion in the previous section, the language of anyons (more precisely, the language of unitary modular tensor category (UMC)) is useful in discussing RCFTs, since they share the same algebraic structure. In this section, we will propose a physical setup to demonstrate that the correspondence betweeen scrambling in RCFTs and anyons in $(2+1)$-dimensions is not a formal connection, but has a intrinsic physical reason, coming from the bulk-boundary correspondence of $(2+1)$-d TOS.



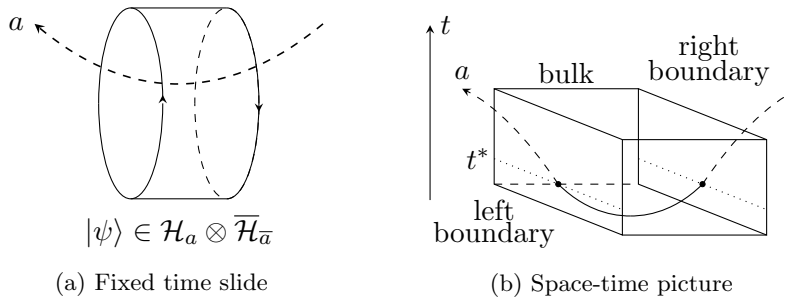
\begin{figure}[h]
\center
\subfloat[Fixed time slide]{
\begin{tikzpicture}[scale=1.2,baseline={(current bounding box.center)}]
\draw[>=stealth,>-] (0pt,0pt) arc (0:180: 10pt and 30pt);
\draw (0pt,0pt) arc (0:-180: 10pt and 30pt);
\draw[dashed] (30pt,0pt) arc (0:180: 10pt and 30pt);
\draw[dashed, >=stealth,>-] (30pt,0pt) arc (0:-180: 10pt and 30pt);
\draw(20pt,-30pt) arc (-90:90: 10pt and 30pt);
\draw (-10pt,-30pt)--(20pt,-30pt);
\draw (-10pt,30pt)--(20pt,30pt);
\draw[line width=0.6pt, dashed, >=stealth, ->] (50pt,25pt)..controls (20pt,0pt) and (-10pt,0pt).. (-40pt,25pt) node[left]{$a$};
\node at (5pt,-40pt) {$|\psi \rangle \in \mathcal{H}_a \otimes \overline{\mathcal{H}}_{{\bar{a}}}$};
\end{tikzpicture}
\label{subfig:fixed time slide of bulk-boundary}
}\hspace{30pt}\subfloat[Space-time picture]{
\begin{tikzpicture}[scale=1.2,baseline={(current bounding box.center)}]
\draw[yslant=-0.4] (0pt,0pt) rectangle (40pt,30pt);
\draw[xshift=45pt, yslant=-0.4] (0pt,0pt) rectangle (40pt,30pt);
\draw (42.5pt,-10pt) ..controls (40pt,-10pt) and (30pt,-10pt).. (20pt,0pt);
\draw (42.5pt,-10pt) ..controls (45pt,-10pt) and (55pt,-10pt).. (65pt,0pt);
\draw[->,dashed,>=stealth] (20pt,0pt)..controls (10pt,15pt) and (0pt,25pt)..(-10pt,30pt) node[above]{$a$}; 
\draw[dashed,>=stealth] (65pt,0pt)..controls (75pt,15pt) and (85pt,25pt).. (95pt,30pt);
\filldraw (20pt,0pt) circle (0.8pt);
\draw[dotted] (0pt,8pt)--(40pt,-8pt);
\draw[dotted,xshift=45pt](0pt,8pt)--(40pt,-8pt);
\node at (-5pt,8pt){$t^*$};
\filldraw (65pt,0pt) circle (0.8pt);
\draw[dashed] (0pt,0pt)--(45pt,0pt);
\draw (0pt,30pt)--(45pt,30pt);
\draw (40pt,-16pt)--(85pt,-16pt);
\draw (40pt,14pt)--(85pt,14pt);
\node at (22.5pt,35pt){bulk};
\node at (0pt,-8pt){left};
\draw [>=stealth,->] (-20pt,-5pt)--(-20pt,50pt) node[right]{$t$};
\node at (0pt,-16pt){boundary};
\node at (67pt,43pt){right};
\node at (67pt,35pt){boundary};
\end{tikzpicture}
\label{subfig:spacetime of bulk-boundary}
}
\caption{The bulk-boundary correspondence. (a) At a fixed time $t^*$, the CFT state at two boundaries belong to sector $\mathcal{H}_a\otimes \mathcal{H}_{\bar{a}}$; (b) The spacetime picture for a pair of anyons $a,\bar{a}$ created and passed through the boundary at fixed time $t^*$. The state was in identity sector before $t^*$: $\psi(t<t^*)\in \mathcal{H}_1 \otimes \overline{\mathcal{H}}_1$, and shifted to sector $\mathcal{H}_a \otimes \overline{\mathcal{H}}_{\bar{a}}$ after time $t^*$.}
\label{fig:bulk-boundary correspondence}
\end{figure}

Chiral part of RCFT (i.e. the holomorphic sectors, or the anti-holomorphic sectors) can be realized as the edge theory of $(2+1)$-d chiral topological order\cite{witten1989quantum,moore1990lectures}, such as fractional quantum Hall (FQH) states\cite{moore1991nonabelions,read1999beyond}. In this bulk-boundary correspondence, a non-chiral RCFT can be viewed as the low energy effective theory of a strip of $(2+1)$-d chiral topological order, as is shown in figure~\ref{fig:bulk-boundary correspondence}. In such a strip, the edge states on the two boundaries are described by the holomorphic and anti-holomorphic sectors of a RCFT, and the bulk topological order is described by the corresponding UMC. There is a one-to-one correspondence between the labels of primary fields in RCFT and the labels of anyon types in the bulk. When the bulk has no anyon (or has anyons that fuse to identity, i.e., have zero total anyonic charge), the topological sector of the two boundaries must be conjugate of each other, so that the low energy Hilbert space of this system is the same as that of a diagonal RCFT:
\begin{eqnarray}
\mathcal{H}_{total}=\bigoplus_{a} \mathcal{H}_a \otimes \overline{\mathcal{H}}_{\bar{a}}\label{Hilbertspace}
\end{eqnarray} 

One can also consider the physical process in the $(2+1)$-d strip corresponding to acting operators in the RCFT in the spacetime picture, see figure~\ref{subfig:spacetime of bulk-boundary}. The insertion of a primary operator of family $(a,\bar{a})$ at $(x,t^*)$ on the boundary corresponds to creating a pair of anyons $(a,\bar{a})$ at an earlier time and pass them through the boundary at spacetime point $(x,t^*)$. To compute two point functions, one need to get them back at a later time and annihilate them in the bulk. 

\begin{figure}[h]
\center
\begin{tikzpicture}[scale=2,baseline={(current bounding box.center)}]
\draw[yslant=-0.4] (0pt,0pt) rectangle (40pt,60pt);
\draw[xshift=45pt, yslant=-0.4] (0pt,0pt) rectangle (40pt,60pt);
\draw [xshift=10pt,>=stealth,->] (-20pt,-5pt)--(-20pt,20pt) node[left]{$t$};
\draw [xshift=10pt,>=stealth,->] (-20pt,-5pt)--(10pt,-17pt) node[below]{$x$};
\node at (35pt,-8pt){L};
\node at (80pt,-8pt){R};
\filldraw (20pt,0pt) circle (0.8pt) node[left]{$V_L$};
\filldraw (20pt,30pt) circle (0.8pt) node[left]{$W_L$};
\filldraw (23.5pt,41.2pt) circle (0.8pt) node[right]{$W_L^\dagger$};
\filldraw (30pt,32pt) circle (0.8pt) node[right]{$V_L^\dagger$};
\filldraw (65pt,0pt) circle (0.8pt) node[left]{$V_R$};
\filldraw (55pt,40pt) circle (0.8pt) node[left]{$V^\dagger_R$};
\filldraw (65pt,30pt) circle (0.8pt)  node[right]{$W_R$};
\filldraw (61.5pt,44pt) circle (0.8pt) node[right]{$W_R^\dagger$};
\draw[dashed,->,>=stealth](20pt,30pt)--(23.5pt,41.2pt);
\draw[dotted] (0pt,38pt)--(40pt,22pt);
\node[left] at (0pt,38pt){ $t$};
\node[left] at (0pt,50.4pt){$t+2c$};
\draw[dotted] (0pt,50.4pt)--(40pt,34.4pt);
\draw[dashed,->,>=stealth] (20pt,0pt)--(30pt,32pt);
\draw[dashed,->,>=stealth](65pt,30pt)--(61.5pt,44pt);
\draw[dotted] (45pt,38pt)--(85pt,22pt);
\draw[dotted] (45pt,50.4pt)--(85pt,34.4pt);
\draw[dashed,->,>=stealth] (65pt,0pt)--(55pt,40pt);
\end{tikzpicture}
\caption{OTOC is mapped to a time-ordered four-point function on the two boundaries of a strip, by shifting along the light cones (see equation~(\ref{eqn:OTO after shifting}). }
\label{fig:oto correlation function after shifting}
\end{figure}
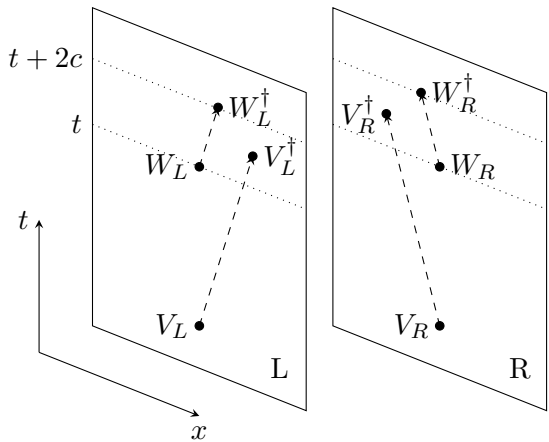

For four point functions, the procedure is similar. However, we will show in details that there is a nontrivial linking structure in the bulk picture for OTOC. In our two-side setting, one can split each physical operator into left-moving and right-moving parts, acting on the left and right boundaries of the strip: $\mathcal{O}(x,t)=\mathcal{O}_L(x,t) \mathcal{O}_R(x,t)$. The advantage of such a decomposition is that we are allowed to move the chiral operators along light cones freely: $\mathcal{O}_L(x,t) = \mathcal{O}_L(x-c,t-c)$ and $\mathcal{O}_R(x,t) = \mathcal{O}_R(x+c,t-c) $ where $c$ is an arbitrary constant. Therefore, we can use this freedom to map the OTOC to a time-ordered four-point function:
\begin{align}
f(t)=\langle W(t)^\dagger V^\dagger W(t) V \rangle_\beta
=&\langle W^\dagger_L(2c,t+2c) V^\dagger_L(t+c,t+c) W_L(0,t) V_L(0,0)\cdot \nonumber \\
& W^\dagger_R(-2c,t+2c) V^\dagger_R(-t-c,t+c) W_R(0,t) V_R(0,0) \rangle_\beta \label{eqn:OTO after shifting}
\end{align}
with $c>0$. After this shift, we can compute the OTOC in time ordered way, as shown in figure~\ref{fig:oto correlation function after shifting}.

\begin{figure}[h]
\center
\subfloat[]{
\begin{tikzpicture}[scale=1.2,baseline={(current bounding box.center)}]
\draw[yslant=-0.4] (0pt,0pt) rectangle (40pt,60pt);
\draw[xshift=45pt, yslant=-0.4] (0pt,0pt) rectangle (40pt,60pt);
\draw[red,line width=0.8pt](65pt,30pt)..controls(45pt,20pt) and (35pt,20pt) ..(20pt,30pt);
\draw[densely dotted, red, line width=0.8pt] (20pt,30pt)..controls (16pt,33pt) and (16pt,38pt)..(23.5pt,41.2pt);

\draw [red,line width=0.8pt](23.5pt,41.2pt)..controls(40pt,50pt) and (50pt,50pt) ..(61.5pt,44pt);
\draw[line width=0.8pt] (65pt,0pt) ..controls (50pt,-10pt) and (30pt,-10pt).. (20pt,0pt);
\draw[line width=0.8pt] (55pt,40pt)..controls (50pt,45pt) and (38pt,40pt) ..(30pt,32pt);
\draw[line width=7pt, white](20pt,0pt)..controls (10pt, 10pt) and (20pt,20pt)..(30pt,32pt);
\draw[line width=0.8pt, densely dotted](20pt,0pt)..controls (10pt, 10pt) and (20pt,20pt)..(30pt,32pt);
\draw[line width=0.8pt, densely dotted](55pt,40pt)..controls (75pt,30pt) and (85pt,10pt)..(65pt,0pt);

\draw[white, line width=7pt] (61.5pt,44pt)..controls (75pt,38pt) and (75pt,34pt)..(65pt,30pt);
\draw[densely dotted, red, line width=0.8pt] (61.5pt,44pt)..controls (75pt,38pt) and (75pt,34pt)..(65pt,30pt);
\draw[dashed] (0pt,0pt)--(45pt,0pt);
\draw (0pt,60pt)--(45pt,60pt);
\draw (40pt,-16pt)--(85pt,-16pt);
\draw (40pt,44pt)--(85pt,44pt);
\filldraw (20pt,0pt) circle (0.8pt);
\filldraw (20pt,30pt) circle (0.8pt);
\filldraw (23.5pt,41.2pt) circle (0.8pt);
\filldraw (65pt,0pt) circle (0.8pt);
\filldraw (55pt,40pt) circle (0.8pt);
\filldraw (30pt,32pt) circle (0.8pt);
\filldraw (65pt,30pt) circle (0.8pt);
\filldraw (61.5pt,44pt) circle (0.8pt);
\draw[dotted] (0pt,38pt)--(40pt,22pt);
\draw[dotted] (0pt,50.4pt)--(40pt,34.4pt);
\draw[dotted] (45pt,38pt)--(85pt,22pt);
\draw[dotted] (45pt,50.4pt)--(85pt,34.4pt);
\end{tikzpicture}
\label{fig:bulk OTOC}
}
\hspace{35pt}
\subfloat[]{
\begin{tikzpicture}[scale=1.2, baseline={(current bounding box.center)}]
\draw[xshift=75pt, yslant=-0.4] (0pt,0pt) rectangle (40pt,60pt);
\draw[line width=5pt, white, xshift=45pt,yslant=-0.4] (0pt,0pt) rectangle (40pt,60pt);
\draw[densely dotted,xshift=45pt, yslant=-0.4] (0pt,0pt) rectangle (40pt,60pt);
\draw[red,line width=0.8pt](65pt,30pt)..controls(45pt,20pt) and (35pt,20pt) ..(20pt,30pt);
\draw[red, line width=0.8pt] (20pt,30pt)..controls (16pt,33pt) and (16pt,38pt)..(23.5pt,41.2pt);
\draw[line width=0.8pt] (65pt,0pt) ..controls (50pt,-10pt) and (30pt,-10pt).. (20pt,0pt);
\draw[line width=0.8pt] (55pt,40pt)..controls (50pt,45pt) and (38pt,40pt) ..(30pt,32pt);
\draw[line width=5pt, white](20pt,0pt)..controls (10pt, 10pt) and (20pt,20pt)..(30pt,32pt);
\draw[line width=0.8pt](20pt,0pt)..controls (10pt, 10pt) and (20pt,20pt)..(30pt,32pt);
\draw[line width=0.8pt](55pt,40pt)..controls (75pt,30pt) and (85pt,10pt)..(65pt,0pt);
\draw[white, line width=5pt] (61.5pt,44pt)..controls (75pt,38pt) and (75pt,34pt)..(65pt,30pt);
\draw[red, line width=0.8pt] (61.5pt,44pt)..controls (75pt,38pt) and (75pt,34pt)..(65pt,30pt);
\draw [red,line width=0.8pt](23.5pt,41.2pt)..controls(40pt,50pt) and (50pt,50pt) ..(61.5pt,44pt);
\filldraw (20pt,0pt) circle (1.0pt);
\filldraw (20pt,30pt) circle (1.0pt);
\filldraw (23.5pt,41.2pt) circle (1.0pt);
\filldraw (65pt,0pt) circle (1.0pt);
\filldraw (55pt,40pt) circle (1.0pt);
\filldraw (30pt,32pt) circle (1.0pt);
\filldraw (65pt,30pt) circle (1.0pt);
\filldraw (61.5pt,44pt) circle (1.0pt);
\draw[yslant=-0.4,densely dotted] (0pt,0pt) rectangle (40pt,60pt);
\draw[line width=5pt, white, xshift=-30pt, yslant=-0.4] (0pt,0pt) rectangle (40pt,60pt);
\draw[xshift=-30pt, yslant=-0.4] (0pt,0pt) rectangle (40pt,60pt);
\draw[dashed] (-30pt,0pt)--(75pt,0pt);
\draw (10pt,-16pt)--(115pt,-16pt);
\draw (-30pt,60pt)--(75pt,60pt);
\draw (10pt,44pt)--(115pt,44pt);
\draw[->,>=stealth] (-30pt,70pt)--(0pt,70pt);
\draw[->,>=stealth] (75pt,70pt)--(45pt,70pt);
\end{tikzpicture}
\label{fig: bulk boundary:move the boundary}
}
\caption{(a) Anyon world-lines corresponding to the OTOC. Black lines (solid for interior, dotted for exterior) are for anyon label $b$, and red lines are for anyon label $a$. The insertion point at the boundary is determined by equation~(\ref{eqn:OTO after shifting}) and figure~\ref{fig:oto correlation function after shifting}. (b) The interpolation between a completely bulk anyon braiding procedure and a boundary OTOC. Starting with a pair of linked anyon world-lines inside the bulk, one can move both boundaries inwards to intersect with the world-lines. Such world lines will induce insertion of operators in the families $a,b$. By proper arrangement of initial configurations of the linked world lines, one obtains the OTOC in subfigure (a). 
}
\end{figure}
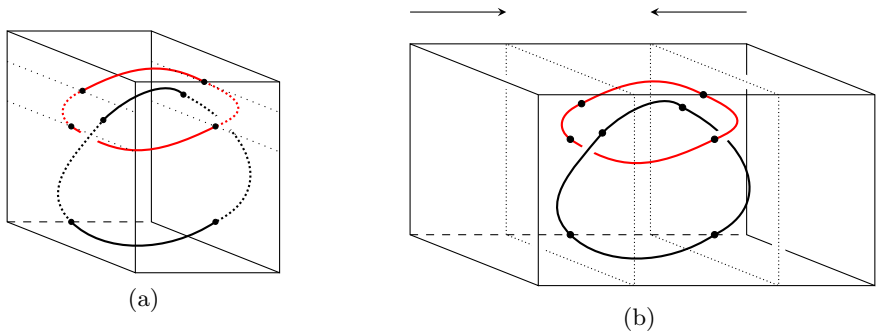

From the viewpoint of bulk-boundary correspondence (figure~\ref{fig:bulk OTOC}), the insertions of operators $V[b],V^\dagger[\bar{b}]$ on two boundaries corresponds to creating a pair of anyons labeled by $b$, $\bar{b}$ and passing through the boundary at insertion points of $V_L,V_R$, then pull the anyons back at the insertion points of $V^\dagger_L, V^\dagger_R$. Similar discussion applies to operators $W[a],W^\dagger[\bar{a}]$. Remarkably, this configuration of operators on the boundaries induce a nontrivial linking of anyon worldlines in the bulk, as is shown in figure~\ref{fig:bulk OTOC}. One can further compute the linked world lines in the anyon theory, which gives $\mathcal{D}s_{ab}^*$. The computation is under the assumption that bulk energy gap of topological quantum matter is much larger than the temperature $1/\beta$ such that we can ignore the actual dynamics of the bulk particles and only focus on the topological content.

Alternatively, one can interpret the bulk-boundary correspondence here by starting with a pair of world-lines for anyons $a$ and $b$, see figure~\ref{fig: bulk boundary:move the boundary}, and then imagine to move the boundaries inwards. The boundaries will finally intersect with the world lines, which induce insertions of operators on the boundaries. If we start with the linked configuration as shown in figure~\ref{fig: bulk boundary:move the boundary}, we are able to arrange the crossing points such that they correspond to the OTOC in equation~(\ref{eqn:OTO after shifting}).

As a comparison, one can run the same procedure for ordinary ordered correlation function $g(t)$, which can be split and shifted to time ordered as well:
\begin{align}
g(t)=\langle V^\dagger W^\dagger(t)  W(t) V \rangle_\beta 
=& \langle V_L^\dagger(t+c,t+c) W^\dagger_L(0,t)  W_L(0,t) V_L(0,0)\cdot \nonumber \\
&V_R^\dagger(-t-c,t+c) W^\dagger_R(0,t)  W_R(0,t) V_R(0,0)\rangle_\beta 
\label{eqn: ordinary ordered four point function g(t)}
\end{align}

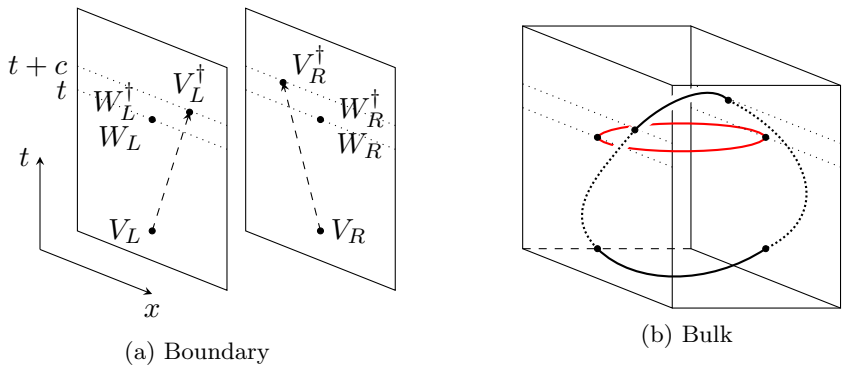
\begin{figure}[h]
\center
\subfloat[Boundary]{
\begin{tikzpicture}[scale=1.4,baseline={(current bounding box.center)}]
\draw[yslant=-0.4] (0pt,0pt) rectangle (40pt,60pt);
\draw[xshift=45pt, yslant=-0.4] (0pt,0pt) rectangle (40pt,60pt);
\draw[dotted] (0pt,38pt)--(40pt,22pt);
\node[left] at (0pt,38pt){$t$};
\node[left] at (0pt,44.4pt){$t+c$};
\draw[dotted] (0pt,44.4pt)--(40pt,28.4pt);
\draw[dashed,->,>=stealth] (20pt,0pt)--(30pt,32pt);
\draw[dotted] (45pt,38pt)--(85pt,22pt);
\draw[dotted] (45pt,44.4pt)--(85pt,28.4pt);
\draw[dashed,->,>=stealth] (65pt,0pt)--(55pt,40pt);
\filldraw (55pt,40pt) circle (0.8pt);
\filldraw (30pt,32pt) circle (0.8pt) node[above]{$V_L^\dagger$};
\filldraw (20pt,0pt) circle (0.8pt) node[left]{$V_L$};
\filldraw (20pt,30pt) circle (0.8pt);
\node at (62pt,45pt) {$V_R^\dagger$};
\node at (10pt,35pt){$W_L^\dagger$};
\node at (11.5pt,25pt){$W_L$};
\filldraw (65pt,30pt) circle (0.8pt);
\node at (76pt,33pt){$W_R^\dagger$};
\node at (75.5pt,23pt){$W_R$};
\filldraw (65pt,0pt) circle (0.8pt)node[right]{$V_R$};

\draw [>=stealth,->] (-10pt,-5pt)--(-10pt,20pt) node[left]{$t$};
\draw [>=stealth,->] (-10pt,-5pt)--(20pt,-17pt) node[below]{$x$};
\end{tikzpicture}
\label{subfig:g(t) boundary}
}
\hspace{40pt}
\subfloat[Bulk]{\begin{tikzpicture}[scale=1.4,baseline={(current bounding box.center)}]
\draw[yslant=-0.4] (0pt,0pt) rectangle (40pt,60pt);
\draw[xshift=45pt, yslant=-0.4] (0pt,0pt) rectangle (40pt,60pt);
\draw[red,line width=0.8pt] (20pt,30pt)..controls (23pt,25pt)and (62pt,25pt)..(65pt,30pt);
\draw[red,line width=0.8pt] (20pt,30pt)..controls (23pt,35pt)and (62pt,35pt)..(65pt,30pt);
\draw[dotted] (0pt,38pt)--(40pt,22pt);
\draw[dotted] (0pt,44.4pt)--(40pt,28.4pt);
\draw[dotted] (45pt,38pt)--(85pt,22pt);
\draw[dotted] (45pt,44.4pt)--(85pt,28.4pt);
\draw[line width=0.8pt] (65pt,0pt) ..controls (50pt,-10pt) and (30pt,-10pt).. (20pt,0pt);
\draw[line width=5pt,white] (55pt,40pt)..controls (50pt,45pt) and (38pt,40pt) ..(30pt,32pt);
\draw[line width=0.8pt] (55pt,40pt)..controls (50pt,45pt) and (38pt,40pt) ..(30pt,32pt);
\draw[line width=5pt, white](20pt,0pt)..controls (10pt, 10pt) and (20pt,20pt)..(30pt,32pt);
\draw[line width=0.8pt, densely dotted](20pt,0pt)..controls (10pt, 10pt) and (20pt,20pt)..(30pt,32pt);
\draw[line width=0.8pt, densely dotted](55pt,40pt)..controls (75pt,30pt) and (85pt,10pt)..(65pt,0pt);
\filldraw (20pt,0pt) circle (0.8pt);
\filldraw (20pt,30pt) circle (0.8pt);
\filldraw (65pt,30pt) circle (0.8pt);
\filldraw (65pt,0pt) circle (0.8pt);
\filldraw (55pt,40pt) circle (0.8pt);
\filldraw (30pt,32pt) circle (0.8pt);
\draw[dashed] (0pt,0pt)--(45pt,0pt);
\draw (0pt,60pt)--(45pt,60pt);
\draw (40pt,-16pt)--(85pt,-16pt);
\draw (40pt,44pt)--(85pt,44pt);
\end{tikzpicture}
\label{subfig:g(t) bulk}}
\caption{Bulk-boundary correspondence for ordinary ordered four point function $g(t)$. (a) shows the insertion of operators $V[b],W[a]..$ on two boundaries of a strip of topological quantum matter, we shifts the operator along light cone to reach the position in equation~(\ref{eqn: ordinary ordered four point function g(t)}). (b) shows the corresponding anyon world-lines, red for label $a$ and black for label $b$, such world-line configuration doesn't link, therefore can be deform to two independent loops. (Notice that $W_{L/R}$ and $W^\dagger_{L/R}$ share the same location in figures.)}
\label{fig: bulk-boundary: ordinary correlation}
\end{figure}
The bulk picture for $g(t)$ was shown in figure~\ref{subfig:g(t) bulk}, where we can clearly see that the world lines are unlinked, which therefore, correspond to two independent loops with labels $a$ and $b$. Such loops together have the amplitude $d_a d_b$ according to theory of anyons.

In summary, we have provided a physical interpretation to the correspondence between late-time universal behavior of OTOC in RCFT and fractional statistics of anyons in $(2+1)$-d TOS. The intrinsic reason of this correspondence is the fact that these two seemingly unrelated phenomena---the butterfly effect and fractional statistics---are both consequences of nonlocality in unitary time evolutions. The butterfly effect comes from the propagation of quantum information in the Hilbert space from simple local operators to more and more non-local operators (for related discussion see Ref. \cite{hosur2015chaos}) which makes it more and more difficult to reveal the information locally. Similarly, fractional statistics is only possible because anyons are intrinsically non-local. The braiding of anyons, especially that of non-Abelian anyons, lead to a nontrivial and non-local unitary transformation on the Hilbert space, which can be viewed as a special example of scrambling. 


\section{Out-of-time-ordered-correlators of random operators}

In the previous sections, we have discussed how OTOC of a pair of operators, each in a prefixed conformal family, depends on the algebraic content of the operators. Indeed, by the bulk-boundary correspondence, we show that OTOC in RCFT is related to the braiding of anyons in $(2+1)$-d. In this section, we would like to investigate how OTOC can be used as a diagnostic of RCFTs. For that purpose, we would like to consider OTOC of a generic pair of operators, rather than selecting particular operators by hand. A natural choice is to consider OTOC between two randomly chosen operators. The key question we need to address is what is a natural random ensemble of operators in an RCFT. 

To specify the random ensemble, we need a proper probability distribution for conformal families. Since the Hilbert space of RCFT is a direct sum of that of each conformal family (equation~(\ref{Hilbertspace})), the probability of a random vector in the Hilbert space to be in a given conformal family is proportional to the Hilbert space dimension of that family, i.e., $p_a\propto \operatorname{dim} (\mathcal{H}_a \otimes \overline{\mathcal{H}}_{\bar{a}})$. Naively this is not meaningful, since the Hilbert space dimension of each family is infinite. Nevertheless, the ratio of the Hilbert space dimensions of different sectors is well-defined and is determined by the quantum dimension:
\begin{eqnarray}
\frac{\operatorname{dim} \mathcal{H}_a}{\operatorname{dim} \mathcal{H}_1}= \frac{\chi_a(0)}{\chi_1(0)}= \frac{s_{a1}}{s_{11}}=d_a
\label{eqn:quantum dimension}
\end{eqnarray}
where $\chi_a(\tau)=\Tr_a e^{2\pi i \tau (L_0-c/24)}$ is the character for sector $a$. $\chi_a(0)$ is the high temperature limit of the character.\footnote{$\chi_a(0)$ can be computed by the modular S matrix: $\chi_a(\tau)=\sum_b s_{ab}\chi_b(-1/\tau)$. For $\tau\rightarrow 0i$, $\chi_b(-1/\tau)$ is dominated by the vacuum sector, so that $\chi_a(0)=s_{a1}\chi_1(i\infty)$.\cite{cardy1986operator}} Since only the ratio between different sectors matter for the definition of random ensemble, we can define a regularized probability $p_a=d_a^2/{\mathcal{D}}^2$. When we consider a random operator, it is chosen to be in sector $(a,\bar{a})$ with probability $p_a$. 

It is interesting to note that this probability distribution is also {\it fusion invariant}, 
i.e., $\sum_{a,b}p_a p_b p_{ab\rightarrow c} = p_c$, where $p_{ab\rightarrow c}$ is the probability of fusing $a$ and $b$ to total charge $c$. According to Ref.~\cite{kitaev2006topological}, $p_{ab\rightarrow c}=d_c N_{ab}^c/d_ad_b$. Therefore, 
\begin{eqnarray}
\sum_{a,b}p_ap_bp_{ab\rightarrow c}=\sum_{a,b} \frac{d_a^2 d_b^2}{\mathcal{D}^4}  \frac{d_c N_{ab}^c}{d_ad_b} 
= \frac{\sum_b d_b^2 d_c^2}{\mathcal{D}^4}= p_c
\end{eqnarray}
The fusion invariance further justifies the probability $p_a=d_a^2/\mathcal{D}^2$ as the correct random ensemble. If we consider a random anyon gas in the bulk, and draw two large adjacient regions $A$ and $B$, we expect the anyon type of each region (defined by the fusion of all anyons in that region) to be random, while the same should apply to region $A\cup B$. This is the bulk interpretation why a random distribution $p_a$ that emerges from ergodic motion of anyons should be fusion invariant. 

With this probability distribution, we can compute the random average of the residue value $r[a,b]$:
\begin{eqnarray}
\langle r \rangle :=\sum_{a,b} p_a p_b r[a,b]=\sum_{a,b}\frac{d_a^2 d_b^2}{\mathcal{D}^4} \frac{\mathcal{D} s_{ab}^*}{d_a d_b}
= \sum_{a,b} \frac{s_{1a}s_{1b}s_{ab}^*}{\mathcal{D}}=\frac{1}{\mathcal{D}^2}
\end{eqnarray}
In the second last step, we used the unitarity of S-matrix in the summation. Interestingly, the final result only depends on the total quantum dimension $\mathcal{D}$, which also appeared as the characterization of topological entanglement entropy $\gamma=\log \mathcal{D}$, as was proposed in Ref.~\cite{kitaev2006topological,levin2006detecting}. Therefore we have related a measure of the butterfly effect, the random operator OTOC, with a measure of the topological order, the topological entropy:\footnote{A similar formula appears in Ref. \cite{hosur2015chaos}, where the averaged OTOC is related to the second Renyi entropy of a certain region in the doubled state that represents the time evolution operator. It is possible that these two formula are related, although the relation is not clear to us yet. }
\begin{eqnarray}
\langle r\rangle=e^{-2\gamma}\label{OTOCTE}
\end{eqnarray}

We would like to provide some further analysis to this formula. Firstly, we discussed earlier that for fixed channels, scrambling only occurs for non-Abelian channels, since for Abelian channels $a,b$, $\left|r[a,b]\right|=1$. In contrast, for the random operator case, the average $\langle r\rangle<1$ even for an Abelian theory. This is because even in an Abelian theory, nontrivial phase interference can occur between different conformal families for an operator that is a superposition of different families. Physically, even an Abelian fractional statistics requires fractionalization, which in the $(2+1)$-d language means that even an Abelian anyon is a collective excitation of the system which cannot be created by a local operator. Due to such intrinsic nonlocality in the dynamics of the system, the time evolution of a generic operator looks chaotic, although that of a special operator in a single conformal family does not. In this sense, an Abelian theory is an intermediate case between free (boson or fermion) systems and more chaotic (non-Abelian) RCFTs.

Secondly, the total quantum dimension is, roughly speaking, the size of operator content in an RCFT. equation~(\ref{OTOCTE}) means that an RCFT with more fields is on average more chaotic.  For an RCFT with $N$ conformal families, 
$\mathcal{D}^2=\sum_{a=1}^{N} d_a^2 \geq N$, such that $\langle r \rangle\leq \frac1N$. Moreover, the equal sign can only be reached if and only if the theory is Abelian. In other words, for the same number of conformal families (or anyon types in the $(2+1)$-d language), non-Abelian theories are more chaotic than Abelian theories. 
For example, in the $\operatorname{SU}(2)_k$ WZW model, the number of family is $N=k+1$ and the total quantum dimension is $\mathcal{D}=\frac{\sqrt{(k+2)/2}}{\sin \left( \pi/(k+2) \right)} \sim k^{3/2}$ in large $k$ limit. Therefore, $\langle r \rangle \sim k^{-3}$ in $\operatorname{SU(2)}_k$, while in an Abelian theory with the same $N=k+1$ one would have $\langle r \rangle=(k+1)^{-1}$. This is conceptually consistent with our observation in section~\ref{subsec:example} that most channels in ${\rm SU}(2)_k$ are strongly chaotic in the large $k$ limit.

\section{Conclusion and discussions}

In this article, we studied OTOC in the context of RCFTs, and relate its behavior to the universal algebraic data of RCFT, such as the monodromy matrix and the modular S-matrix. Through the bulk-boundary correspondence of $(2+1)$-d TOS, we pointed out a connection between the OTOC in RCFTs and the fractional statistics in the corresponding TOS. We have shown that the OTOC in an RCFT can be mapped to a time-ordered four-point function which corresponds to a physical process of anyon braiding. In other words, our results point out that the two consequences of ``emergent nonlocality" in (2+1)-d TOS---chaos on the boundary and fractionalization in the bulk---always accompany each other. Furthermore, our result shows that for a fixed channel (meaning fixed conformal families in the boundary, or fixed anyon types in the bulk), scrambling only occurs for non-Abelian anyons, as a consequence of nontrivial interference between different fusion channels. When we consider a pair of random operators rather than operators in a fixed conformal family, the average value of OTOC is determined by the total quantum dimension, so that the ``average degree of chaos" in an RCFT is directly related to the topological entanglement entropy in the bulk.

Besides providing a physical interpretation of the relation between chaos and topological order, the bulk-boundary correspondence we discussed also suggests a potential approach of measuring the OTOC experimentally. In equation~(\ref{eq:monodromy}), we show that the residue value $r[a,b]$ of OTOC is determined by the $(1,1)$ element of monodromy matrix $\widetilde{M}[a,b]_{11}$. Interestingly, the same quantity played an essential role in the interferometry of anyons
\cite{bonderson2007decoherence,overbosch2001inequivalent}, which has been studied extensively in fractional quantum Hall states, both theoretically and experimentally.\cite{chamon1997two,fradkin1998chern,bonderson2006detecting,stern2006proposed,ji2003electronic,willett2009measurement,zhang2009distinct}

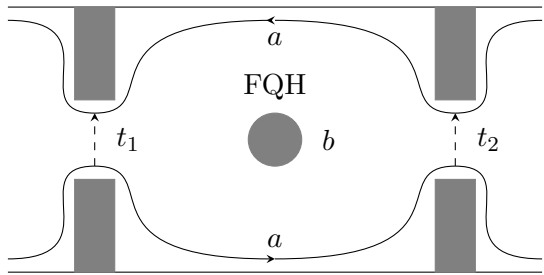
\begin{figure}[h!]
\center
\begin{tikzpicture}[scale=1.0,baseline={(current bounding box.center)}]
\draw (-100pt,-50pt) -- (100pt,-50pt);
\draw (-100pt,50pt) -- (100pt,50pt);
\filldraw[gray] (-75pt,-50pt) rectangle (-60pt,-15pt);
\filldraw[gray] (75pt,-50pt) rectangle (60pt,-15pt);
\filldraw[gray] (-75pt,50pt) rectangle (-60pt,15pt);
\filldraw[gray] (75pt,50pt) rectangle (60pt,15pt);
\draw (-100pt,-45pt)..controls (-60pt,-45pt) and (-95pt,-10pt)..(-67.5pt,-10pt);
\draw[>=stealth,->] (-67.5pt,-10pt)..controls (-40pt,-10pt) and (-75pt,-45pt)..(-0pt,-45pt);
\draw (100pt,-45pt)..controls (60pt,-45pt) and (95pt,-10pt)..(67.5pt,-10pt);
\draw (67.5pt,-10pt)..controls (40pt,-10pt) and (75pt,-45pt)..(-0pt,-45pt);
\draw (-100pt,45pt)..controls (-60pt,45pt) and (-95pt,10pt)..(-67.5pt,10pt);
\draw[>=stealth,-<] (-67.5pt,10pt)..controls (-40pt,10pt) and (-75pt,45pt)..(-0pt,45pt);
\draw (100pt,45pt)..controls (60pt,45pt) and (95pt,10pt)..(67.5pt,10pt);
\draw (67.5pt,10pt)..controls (40pt,10pt) and (75pt,45pt)..(-0pt,45pt);
\filldraw[gray] (0pt,0pt) circle (10pt);
\draw[dashed,>=stealth,->] (-67.5pt,-10pt)--(-67.5pt,10pt);
\draw[dashed,>=stealth,->] (67.5pt,-10pt)--(67.5pt,10pt);
\node at (-55pt,0pt){$t_1$};
\node at (80pt,0pt){$t_2$};
\node at (20pt,0pt){$b$};
\node at (0pt,38pt){$a$};
\node at (0pt,-38pt){$a$};
\node at (0pt,20pt){FQH};
\end{tikzpicture}
\caption{An example of two point-contact interferometry. $t_1$ and $t_2$ are the tunneling amplitude for left and right ``bridges'' respectively. The fractional quantum Hall state occupies the region between the two edges (curves with arrow), with the shaded regions depleted. We denote the anyonic charge on the edge and the central island by $a$ and $b$, respectively.}
\label{fig:twopointcontact}
\end{figure}

To be more precise, consider a typical interferometer of FQH state with two point contacts\cite{chamon1997two}, as is shown in figure~\ref{fig:twopointcontact}. The physical measurable quantity in this setup is the two terminal conductance (with current flowing from left to right) $\sigma_{xx} \propto  |t_1|^2+|t_2|^2 + 2 \operatorname{Re} ( t_1^* t_2 e^{i\alpha_{ab}} \widetilde{M}[a,b]_{11} )$. The phase factor $\exp (i \alpha_{ab})$ includes contributions from the Aharonov-Bohm phase and other dynamical phase factors. Therefore we see that the residual value of normalized OTOC $r[a,b]=\widetilde{M}[a,b]_{11}$ plays an essential role in the conductance oscillation. 
A stronger butterfly effect corresponds to a smaller conductance oscillation. In the extreme case when $\widetilde{M}[a,b]_{11}=0$, no interference can be observed in the conductance. For example such an absence of conductance oscillation has been considered in the filling fraction $\nu=5/2$ state as an evidence of non-Abelian statistics\cite{bonderson2006detecting,stern2006proposed,willett2009measurement}. According to our results, the absence of conductance oscillation can also be viewed as a measure of complete scrambling in the corresponding conformal families of the RCFT describing the boundary. 


In the end, let us make some more speculations about possible generalizations of our results. Chaos is suppressed by emergent conservation laws which constrains the dynamics of the theory. Therefore it is natural to guess that similar results on OTOC can be obtained in non-critical one-dimensional systems with emergent conservation laws, such as integrable models. The integrability of $(1+1)$-d integrable models is described by the Yang-Baxter equation, which has a similar algebraic structure as that in RCFTs and topological order. 
Therefore, it is tempting to guess that OTOC in one dimensional integrable models might also capture universal algebraic properties of the model, which we leave for future study. 

\acknowledgments

We would like to acknowledge helpful discussions with Meng Cheng, Tian Lan and Daniel A. Roberts. This work is supported by the National Science Foundation through the grant No. DMR-1151786. Upon finishing this work, we become aware of the parallel work of Pawel Caputa, Tokiro Numasawa and Alvaro Veliz-Osorio\cite{caputa2016}. We would like to thank Tokiro Numasawa for informing us of their work before posting their paper.

\appendix

\section{Notations and conventions}
\label{appendix: notations and conventions}

In this appendix, we will introduce necessary backgrounds for the notations we used in main text, especially the diagrams. Both subjects of RCFT and anyons have be extensively studied and properly summarized in literature, and we will follow the presentation of lecture note by G. Moore and N. Seiberg~\cite{moore1990lectures} and the reference therein for RCFTs, and Ref.~\cite{kitaev2006anyons} by A. Kitaev for anyon theories. The purpose of this section is to review the diagrammatic conventions for anyon theories, and explain why we are allowed to use them to describe RCFTs.

Let us start with general RCFTs. Such theories have simple analytic properties in physical correlation functions:
\begin{eqnarray}
\langle \phi \phi \ldots \phi \rangle \sim \sum_{i,j=1}^M g_{ij} \mathcal{F}_i(\tau)\overline{\mathcal{F}}_j(\bar{\tau})
\label{physicalcorrelationfunction}
\end{eqnarray}
Where $M$ is a finite number, counting the dimension of the space of conformal blocks. $g_{ij}$ is the coefficient indicating the paring between holomorphic and anti-holomorphic blocks. Holomorphic conformal blocks form a vector space parametrized by moduli $\tau$ (same to anti-holomorphic blocks, by $\bar{\tau}$). The moduli describes the shape of a two dimensional manifold, together with the locations of fields inserted. An alternative geometrical formulation due to D. Friedan and S. Shenker\cite{friedan1987analytic} describes conformal blocks as a vector bundle over moduli space, and the vector bundle is equipped with a fiber-wise metric $g_{ij}$ for physical correlation functions.

To build a connection to the algebraic theory of anyons, it is essential to find the building blocks on both sides. In RCFT side, such object is the conformal blocks associated to a 3-punctured sphere, or intuitively, the {\it chiral half} of three-point functions.\footnote{More abstractly and probably more precisely, one can use the notion of {\it chiral vertex operators}\cite{moore1988polynomial}.} More explicitly, we assume the three punctures were created by insertions of holomorphic fields of family $a,b,c$, and we denote the space of conformal blocks associated to such geometry by $V_{abc}$. The dimension of this space is known as the fusion multiplicity $N_{abc}=\operatorname{dim} V_{abc}$. Such formalism also contains the notion of dual or anti-particles, $\bar{a},\bar{b},\bar{c}$. Then we can lift and lower the indices in convention: e.g., $V_{abc}\simeq V_{c}^{\bar{a}\bar{b}}$. Accordingly, the indices of fusion multiplicity can be lifted or lowered: e.g., $N_c^{\bar{a}\bar{b}}=\operatorname{dim} V_{c}^{\bar{a}\bar{b}}=\operatorname{dim} V_{abc}= N_{abc}$. 

\begin{figure}[htb]
\center
\subfloat[Splitting]{
\begin{tikzpicture}[scale=1]
\draw (0pt,0pt)--(0pt,-30pt);
\draw (0pt,0pt)--(15pt,25.5pt);
\draw (0pt,0pt)--(-15pt,25.5pt);
\node [right] at (5pt,0pt){$\psi \in V_c^{ab}$};
\node at (-15pt, 30pt) {$a$};
\node at (15pt, 31.5pt) {$b$};
\node at (0pt, -35pt) {$c$};
\end{tikzpicture}}
\hspace{30pt}
\subfloat[Fusion]{
\begin{tikzpicture}[scale=1]
\draw (0pt,0pt)--(0pt,30pt);
\draw (0pt,0pt)--(15pt,-25.5pt);
\draw (0pt,0pt)--(-15pt,-25.5pt);
\node[right] at (5pt,0pt){$\psi^\dagger \in V^c_{ab}$};
\node at (-15pt, -30pt) {$a$};
\node at (15pt, -31.5pt) {$b$};
\node at (0pt, 35pt) {$c$};
\end{tikzpicture}}
\caption{Diagrammatic representation of splitting and fusion space.}
\label{fig:splitting and fusion space}
\end{figure}
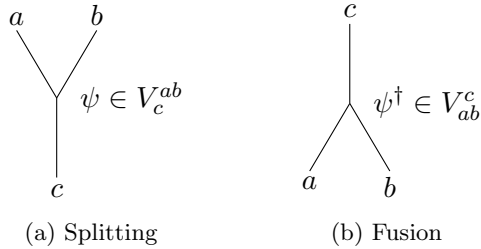

In the algebraic theory of anyons, the parallel notion is the {\it fusion} and {\it splitting} spaces between {\it ``superselection sectors''} $a$, $b$ and $c$ (anyon labels). More concretely, vectors in the splitting space $V_{c}^{ab}$ represent the different ways of splitting $c$ into $a$ and $b$, or equivalent classes of local operators that operate the splitting. Analogously, vectors in the fusion space $V^c_{ab}$ represent the different ways of fusing $a$ and $b$ into $c$. If $\psi \in V_{c}^{ab}$, then its dagger $\psi^\dagger \in V^{c}_{ab}$, see figure~\ref{fig:splitting and fusion space}. $N_{ab}^c=\operatorname{dim} V_c^{ab}$ counts the dimension of splitting/fusion space.

On both sides, such objects can be used to build more complicated spaces: in RCFTs, they are conformal blocks of multiple insertions of operators; in anyons, they are fusion/splitting space of multiple anyons. They also have a set of identical consistent conditions to satisfy, of which the crucial one for the fusion theory is the pentagon equation.

Furthermore, both sides have a notion of braiding: in RCFTs the conformal blocks naturally have a dependence of complex coordinates and the braiding of operators is defined. In anyons, braiding of $a$ and $b$ is an element in $V_{ab}^{ba}$:
\begin{eqnarray}
R_{ab}\in V_{ab}^{ba}: R_{ab}:=
\begin{tikzpicture}[scale=0.7,baseline={(current bounding box.center)}]
\draw (20pt,0pt)..controls (20pt,20pt) and (0pt,20pt)..(0pt,40pt);
\draw[line width=8pt, draw=white] (0pt,0pt)..controls (0pt,20pt) and (20pt,20pt)..(20pt,40pt);
\draw (0pt,0pt)..controls (0pt,20pt) and (20pt,20pt)..(20pt,40pt);
\node at (0pt,-5pt) {$a$};
\node at (20pt,-5pt) {$b$};
\node at (0pt,45pt) {$b$};
\node at (20pt,45pt) {$a$};
\end{tikzpicture}
\label{braiding}
\end{eqnarray}
With the braiding we have more consistent conditions to satisfy, known as the hexagon equations. 

In the end the RCFTs naturally requires modular invariance as a physical constraint, which thus leads to the algebraic structure of {\it modular tensor category} (MTC). In this paper, we further restrict ourselves to {\it diagonal theories}, in which one can choose a proper basis such that the ``metric'' $g_{ij}=\delta_{ij}$. More explicitly, the physical correlation function has a simpler expression in the diagonal basis:
\begin{eqnarray}
\langle \phi \phi \ldots \phi \rangle = \sum_{i=1}^M  \mathcal{F}_i(\tau)\overline{\mathcal{F}}_i(\bar{\tau})\end{eqnarray}
Under this condition, we end up with {\it unitary modular tensor category} (UMC)\cite{moore1989classical}. On the anyon side, after introducing the braiding, the theory is already physically sensible and known as unitary braided fusion category (UBFC). With extra non-degeneracy condition on braiding, we end up with the same algebraic theory: unitary modular tensor category (UMC). We will restrict ourselves in this paper to discuss those theories described by UMC.

Unfortunately, the literature on RCFTs and anyons have different conventions of drawing diagrams and denoting matrices. We will choose the conventions of anyon theories. More explicitly, we follow the ``arrowless'' conventions in Ref.~\cite{kitaev2006anyons}, i.e., the vertical lines should read from bottom to top, and arrows will only be marked on horizontal lines when necessary. For example, in such conventions, the conformal block in equation~(\ref{eqn:conformal block power expansion}) is drawn as a vector in $V_{\bar{a}a\bar{b}b}\simeq V_{b}^{a\bar{a}b}$:
\begin{equation}
\begin{tikzpicture}[scale=1.5,baseline={(current bounding box.center)}]
\draw (0pt,0pt)--(15pt,25pt);
\draw (-15pt,25pt)--(0pt,-0pt);
\draw (-7.5pt,12.5pt)--(0pt,25pt);
\draw (0pt,0pt)--(0pt,-20pt);
\node at (0pt,-25pt){$b(\infty)$};
\node at (-15pt, 30pt){$a(0)$};
\node at (0pt,30pt){$\bar{a}(\eta)$};
\node at (15pt, 30pt){$b(1)$};
\node at (-10pt,5pt){$p$};
\end{tikzpicture}:=\mathcal{F}_p(\eta) \sim \eta^{h_p},\ \eta \sim \exp(-2\pi t/\beta) \ll 1
\end{equation}
we only write the leading term here for the late time $t\gg\beta$.

\section{The Monodromy matrix $\widetilde{M}[a,b]$}
\label{appendix: monodromy}

In this section, we review the algebraic expression of $\widetilde{M}[a,b]$ in terms of F-matrices and R-matrices, and also its diagrammatic expression. Relevant discussions can be found in Ref.~\cite{bonderson2007non}

Operator $\widetilde{M}$ is a linear map: $V^{a\bar{a} b}_b \rightarrow V_b^{a\bar{a}b}$.
\begin{eqnarray}
\widetilde{M}
\begin{tikzpicture}[baseline={(current bounding box.center)}]
\draw (0pt,0pt)--(15pt,25pt);
\draw (-15pt,25pt)--(0pt,-0pt);
\draw (-7.5pt,12.5pt)--(0pt,25pt);
\draw (0pt,0pt)--(0pt,-20pt);
\node at (0pt,-25pt){$b$};
\node at (-15pt, 30pt){$a$};
\node at (0pt,30pt){$\bar{a}$};
\node at (15pt, 30pt){$b$};
\node at (-10pt,5pt){$j$};
\end{tikzpicture}
:=
\begin{tikzpicture}[baseline={(current bounding box.center)}]
\draw (-7.5pt,12.5pt)..controls (-3.75pt,18.75pt) and (3.75pt,6.25pt)..(7.5pt,12.5pt);
\draw[line width=4pt,draw=white] (0pt,0pt)..controls (3.75pt,6.25pt) and (-1.5pt,12.5pt)..(0.75pt,15.75pt);
\draw (0pt,0pt)..controls (3.75pt,6.25pt) and (-1.5pt,12.5pt)..(0.75pt,15.75pt);
\draw (0.75pt,15.75pt)..controls (3pt, 20pt) and (7.5pt,12.5pt)..(15pt,25pt);
\draw [line width=4pt,draw=white] (7.5pt,12.5pt)..controls (9.25pt,19.75pt) and (0.75pt,18.75pt)..(0pt,25pt);
\draw (7.5pt,12.5pt)..controls (9.25pt,19.75pt) and (0.75pt,18.75pt)..(0pt,25pt);
\draw (-15pt,25pt)--(0pt,-0pt);
\draw (0pt,0pt)--(0pt,-20pt);
\node at (0pt,-25pt){$b$};
\node at (-15pt, 30pt){$a$};
\node at (0pt,30pt){$\bar{a}$};
\node at (15pt, 30pt){$b$};
\node at (-10pt,5pt){$j$};
\end{tikzpicture}=
\left( R_{b\bar{a}} R_{\bar{a}b} \right)^{-1}
\begin{tikzpicture}[baseline={(current bounding box.center)}]
\draw (0pt,0pt)--(15pt,25pt);
\draw (-15pt,25pt)--(0pt,-0pt);
\draw (-7.5pt,12.5pt)--(0pt,25pt);
\draw (0pt,0pt)--(0pt,-20pt);
\node at (0pt,-25pt){$b$};
\node at (-15pt, 30pt){$a$};
\node at (0pt,30pt){$\bar{a}$};
\node at (15pt, 30pt){$b$};
\node at (-10pt,5pt){$j$};
\end{tikzpicture}
\end{eqnarray}
Use bases transformation (F-matrix), we can express $\left( R_{b\bar{a}} R_{\bar{a}b} \right)^{-1}$ operator in terms of $R$ and $F$-matrices.
\begin{eqnarray}
& \left( R_{b\bar{a}} R_{\bar{a}b} \right)^{-1}
\begin{tikzpicture}[scale=0.8,baseline={(current bounding box.center)}]
\draw (0pt,0pt)--(15pt,25pt);
\draw (-15pt,25pt)--(0pt,-0pt);
\draw (-7.5pt,12.5pt)--(0pt,25pt);
\draw (0pt,0pt)--(0pt,-20pt);
\node at (0pt,-25pt){$b$};
\node at (-15pt, 30pt){$a$};
\node at (0pt,30pt){$\bar{a}$};
\node at (15pt, 30pt){$b$};
\node at (-10pt,5pt){$j$};
\end{tikzpicture} \nonumber
=\left( R_{b\bar{a}} R_{\bar{a}b} \right)^{-1}\sum_{k} \left[ F_b^{a\bar{a}b}\right]_{jk}
\begin{tikzpicture}[scale=0.8,baseline={(current bounding box.center)}]
\draw (0pt,0pt)--(15pt,25pt);
\draw (-15pt,25pt)--(0pt,-0pt);
\draw (7.5pt,12.5pt)--(0pt,25pt);
\draw (0pt,0pt)--(0pt,-20pt);
\node at (0pt,-25pt){$b$};
\node at (-15pt, 30pt){$a$};
\node at (0pt,30pt){$\bar{a}$};
\node at (15pt, 30pt){$b$};
\node at (8pt,5pt){$k$};
\end{tikzpicture} \nonumber  \\
&=   \sum_{k} \left[ F_b^{a\bar{a}b}\right]_{jk} \left( R_k^{b\bar{a}} R_k^{\bar{a}b} \right)^{-1}
\begin{tikzpicture}[scale=0.8,baseline={(current bounding box.center)}]
\draw (0pt,0pt)--(15pt,25pt);
\draw (-15pt,25pt)--(0pt,-0pt);
\draw (7.5pt,12.5pt)--(0pt,25pt);
\draw (0pt,0pt)--(0pt,-20pt);
\node at (0pt,-25pt){$b$};
\node at (-15pt, 30pt){$a$};
\node at (0pt,30pt){$\bar{a}$};
\node at (15pt, 30pt){$b$};
\node at (8pt,5pt){$k$};
\end{tikzpicture} \nonumber \\
&= \sum_{k,i} \left[ F_b^{a\bar{a} b}\right]_{jk}\left( R_k^{b\bar{a}} R_k^{\bar{a}b} \right)^{-1} \left[ F_b^{a\bar{a} b}\right]^\dagger_{ki} 
\begin{tikzpicture}[scale=0.8,baseline={(current bounding box.center)}]
\draw (0pt,0pt)--(15pt,25pt);
\draw (-15pt,25pt)--(0pt,-0pt);
\draw (-7.5pt,12.5pt)--(0pt,25pt);
\draw (0pt,0pt)--(0pt,-20pt);
\node at (0pt,-25pt){$b$};
\node at (-15pt, 30pt){$a$};
\node at (0pt,30pt){$\bar{a}$};
\node at (15pt, 30pt){$b$};
\node at (-10pt,5pt){$i$};
\end{tikzpicture}
\end{eqnarray}
In terms of matrix elements:
\begin{eqnarray}
\widetilde{M}[a,b]_{ij}=\sum_{k} \left[ F_b^{a\bar{a} b}\right]_{jk} \left( R_k^{b\bar{a}} R_k^{\bar{a}b} \right)^{-1}\left[ F_b^{a\bar{a} b}\right]^\dagger_{ki} 
\end{eqnarray}
We should clarify again to avoid confusion that $\widetilde{M}[a,b]_{ij}$ is in general a matrix itself. The space of the conformal block with intermediate channel label ``$i$'' (or $j,k$) in the equation is in general $N_{a\bar{a}}^i N_{b\bar{b}}^i$ dimensional, which could be greater than $1$. Therefore, labels $i,j,k$ should be read as labels for space in general. However, the ``monodromy'' scalar $\widetilde{M}[a,b]_{11}$ (denoted as $M_{ab}$ in anyon interferometry literature) is indeed a scalar: $N_{a\bar{a}}^1 N_{b\bar{b}}^1=1$ by axioms in both RCFTs\cite{moore1989classical} and anyons\cite{kitaev2006anyons}.

It is also convenient to rewrite matrix $\left( R^{b\bar{a}}_k R^{\bar{a}b}_k \right)^{-1}$ in terms of the {\it topological spin} $\theta_a$, which is related to conventional spin $s_a$ by $\theta_a=\exp(2\pi i s_a)$ when the latter is defined. $\left( R^{b\bar{a}}_k R^{\bar{a}b}_k \right)^{-1}=\frac{\theta_a \theta_b}{\theta_k}$, so that
\begin{eqnarray}
\widetilde{M}[a,b]_{ij}=\sum_{k} \left[ F_b^{a\bar{a} b}\right]_{jk} \left( \frac{\theta_a \theta_b }{\theta_k} \right)\left[ F_b^{a\bar{a} b}\right]^\dagger_{ki} 
\end{eqnarray}

The diagrammatic expression of this equation is easier to memorize. We use the inner product to single out the matrix element $\widetilde{M}[a,b]_{ij}=\frac{\QTr \left( \psi^\dagger_i \widetilde{M} \psi_j \right) }{ \sqrt{ \QTr \left( \psi^\dagger_i \psi_i \right) \QTr \left(\psi^\dagger_j \psi_j \right) } }$, where denominator (normalization):
\begin{eqnarray}
\QTr (\psi^\dagger_i \psi_i ) = \begin{tikzpicture}[baseline={(current bounding box.center)}]
\draw (0pt,-20pt)--(10pt,0pt);
\draw (-5pt,-10pt)--(0pt,0pt);
\draw (-5pt,-10pt)--(-10pt,0pt);
\draw (-5pt,-10pt)--(0pt,-20pt);
\draw (0pt,-20pt)--(0pt,-25pt);
\draw (0pt,20pt)--(10pt,0pt);
\draw (-5pt,10pt)--(0pt,0pt);
\draw (-5pt,10pt)--(-10pt,0pt);
\draw (-5pt,10pt)--(0pt,20pt);
\draw (0pt,20pt)--(0pt,25pt);
\draw (0pt,25pt).. controls  (0pt,35pt) and (20pt,35pt)..(20pt,0pt)..controls (20pt,-35pt) and (0pt,-35pt).. (0pt,-25pt);
\node at (-15pt,0pt){$a$};
\node at (15pt,0pt){$b$};
\node at (-7pt,-17pt){$i$};
\node at (-7pt,17pt){$i^\dagger$};
\end{tikzpicture} =d_a d_b=
\QTr (\psi^\dagger_j  \psi_j)
\end{eqnarray}
and numerator:
\begin{eqnarray}
\QTr\left(\psi^\dagger_i  \widetilde{M} \psi_j \right) &=&\QTr \left[ \left(
\begin{tikzpicture}[scale=0.8,baseline={(current bounding box.center)}]
\draw (0pt,0pt)--(15pt,25pt);
\draw (-15pt,25pt)--(0pt,-0pt);
\draw (-7.5pt,12.5pt)--(0pt,25pt);
\draw (0pt,0pt)--(0pt,-20pt);
\node at (0pt,-25pt){$b$};
\node at (-15pt, 30pt){$a$};
\node at (0pt,30pt){$\bar{a}$};
\node at (15pt, 30pt){$b$};
\node at (-10pt,5pt){$i$};
\end{tikzpicture}   \right)^\dagger
\cdot
\begin{tikzpicture}[scale=0.8,baseline={(current bounding box.center)}]
\draw (-7.5pt,12.5pt)..controls (-3.75pt,18.75pt) and (3.75pt,6.25pt)..(7.5pt,12.5pt);
\draw[line width=4pt,draw=white] (0pt,0pt)..controls (3.75pt,6.25pt) and (-1.5pt,12.5pt)..(0.75pt,15.75pt);
\draw (0pt,0pt)..controls (3.75pt,6.25pt) and (-1.5pt,12.5pt)..(0.75pt,15.75pt);
\draw (0.75pt,15.75pt)..controls (3pt, 20pt) and (7.5pt,12.5pt)..(15pt,25pt);
\draw [line width=4pt,draw=white] (7.5pt,12.5pt)..controls (9.25pt,19.75pt) and (0.75pt,18.75pt)..(0pt,25pt);
\draw (7.5pt,12.5pt)..controls (9.25pt,19.75pt) and (0.75pt,18.75pt)..(0pt,25pt);
\draw (-15pt,25pt)--(0pt,-0pt);
\draw (0pt,0pt)--(0pt,-20pt);
\node at (0pt,-25pt){$b$};
\node at (-15pt, 30pt){$a$};
\node at (0pt,30pt){$\bar{a}$};
\node at (15pt, 30pt){$b$};
\node at (-10pt,5pt){$j$};
\end{tikzpicture}
\right]=\begin{tikzpicture}[scale=0.8,baseline={(current bounding box.center)}]
\draw (20pt,0pt) circle (20pt);
\draw[line width=6pt, draw=white] (0pt,0pt) circle (20pt);
\draw (0pt,0pt) circle (20pt);
\draw[line width=6pt, draw=white]  (0pt,0pt) arc (-180:-90:20pt);
\draw (0pt,0pt) arc (-180:-90:20pt);
\node at (-15pt,0pt){$a$};
\node at (5pt,0pt){$b$};
\draw[>=stealth,->]  (20pt,-20pt)..controls (15pt,-30pt) and (5pt,-30pt)..(0pt,-20pt);
\draw[>=stealth,->] (0pt,20pt)..controls (5pt,30pt) and (15pt,30pt)..(20pt,20pt);
\node at (10pt,-35pt){$j$};
\node at (10pt,35pt){$i$};
\end{tikzpicture}
\end{eqnarray}
We add back the arrows for horizontal lines to avoid confusion. Therefore, the final diagrammatic representation for element $\tilde{M}[a,b]_{ij}$ is
\begin{eqnarray}
\widetilde{M}[a,b]_{ij}=\frac{1}{d_a d_b}
\begin{tikzpicture}[scale=0.8,baseline={(current bounding box.center)}]
\draw (20pt,0pt) circle (20pt);
\draw[line width=6pt, draw=white] (0pt,0pt) circle (20pt);
\draw (0pt,0pt) circle (20pt);
\draw[line width=6pt, draw=white]  (0pt,0pt) arc (-180:-90:20pt);
\draw (0pt,0pt) arc (-180:-90:20pt);
\node at (-15pt,0pt){$a$};
\node at (5pt,0pt){$b$};
\draw[>=stealth,->]  (20pt,-20pt)..controls (15pt,-30pt) and (5pt,-30pt)..(0pt,-20pt);
\draw[>=stealth,->] (0pt,20pt)..controls (5pt,30pt) and (15pt,30pt)..(20pt,20pt);
\node at (10pt,-35pt){$j$};
\node at (10pt,35pt){$i$};
\end{tikzpicture}
\end{eqnarray}
In particular, the $(1,1)$ element has a nice expression in terms of the modular S-matrix: 
\begin{eqnarray}
\widetilde{M}[a,b]_{11}=\frac{1}{d_a d_b} \ 
\begin{tikzpicture}[scale=0.7,baseline={(current bounding box.center)}]
\draw (20pt,0pt) circle (20pt);
\draw[line width=6pt, draw=white] (0pt,0pt) circle (20pt);
\draw (0pt,0pt) circle (20pt);
\draw[line width=6pt, draw=white]  (0pt,0pt) arc (-180:-90:20pt);
\draw (0pt,0pt) arc (-180:-90:20pt);
\node at (-15pt,0pt){$a$};
\node at (5pt,0pt){$b$};
\end{tikzpicture}=\frac{\mathcal{D}s_{ab}^*}{d_a d_b}
\end{eqnarray}

More generally, if one introduce the ``generalized'' S-matrix\cite{moore1989classical,kitaev2006anyons} $S_z\in \operatorname{Aut}\left( \mathcal{L}_z \right)$, where $\mathcal{L}_z = \bigoplus_b V_b^{bz}$ is the space associated to torus with one puncture of label $z$, the $(1z)$ or $(z1)$ element also has simpler algebraic expression in terms of the modular data:
\begin{eqnarray}
\widetilde{M}[a,b]_{1z}=\frac{1}{d_a d_b}
\begin{tikzpicture}[scale=0.7,baseline={(current bounding box.center)}]
\draw (20pt,0pt) circle (20pt);
\draw[line width=6pt, draw=white] (0pt,0pt) circle (20pt);
\draw (0pt,0pt) circle (20pt);
\draw[line width=6pt, draw=white]  (0pt,0pt) arc (-180:-90:20pt);
\draw (0pt,0pt) arc (-180:-90:20pt);
\node at (-15pt,0pt){$a$};
\node at (5pt,0pt){$b$};
\draw[>=stealth,->]  (20pt,-20pt)..controls (15pt,-30pt) and (5pt,-30pt)..(0pt,-20pt);
\node at (10pt,-35pt){$z$};
\end{tikzpicture}=\frac{\mathcal{D} s_{z,a\bar{b}}}{d_a d_b};\qquad
\widetilde{M}[a,b]_{z1}=\frac{1}{d_a d_b}
\begin{tikzpicture}[scale=0.7,baseline={(current bounding box.center)}]
\draw (20pt,0pt) circle (20pt);
\draw[line width=6pt, draw=white] (0pt,0pt) circle (20pt);
\draw (0pt,0pt) circle (20pt);
\draw[line width=6pt, draw=white]  (0pt,0pt) arc (-180:-90:20pt);
\draw (0pt,0pt) arc (-180:-90:20pt);
\node at (-15pt,0pt){$a$};
\node at (5pt,0pt){$b$};
\draw[>=stealth,->] (0pt,20pt)..controls (5pt,30pt) and (15pt,30pt)..(20pt,20pt);
\node at (10pt,35pt){$z$};
\end{tikzpicture}=\frac{\mathcal{D} s_{\bar{z},a\bar{b}}}{d_a d_b}
\end{eqnarray}
If we take $z=1$, $s_{z,a\bar{b}}=s_{\bar{z},a\bar{b}}=s_{a\bar{b}}=s_{ab}^*$ goes back to the familiar S-matrix. (To avoid confusion, we remark here that $s_{z,ab}$ we defined via diagram is different from Ref.~\cite{kitaev2006anyons} by a factor of $\sqrt{d_z}$ due to the different conventions for diagram normalization. Our $s_{z,ab}$ has normalization: $\sum_b s_{z,ab} s^*_{z,bc} = d_z \delta_{ac}$.)

\section{The residue value $r$ in ${\rm SU(N)}_2$ WZW models}
\label{appendix: detailed WZW}

In this section, we study the residue value $r$ in ${\rm SU(N)}$ WZW models at level $2$, which are related to the $\operatorname{SU}(2)$ level $N$ models we study in the main text by the level-rank duality.

\begin{figure}[h]
\center
\subfloat[]{
\begin{tikzpicture}[baseline={(current bounding box.center)}]
\draw (0pt,0pt) rectangle (20pt,20pt);
\draw[xshift=20pt] (0pt,0pt) rectangle (20pt,20pt);
\draw[yshift=-20pt](0pt,0pt) rectangle (20pt,20pt);
\draw[yshift=-40pt] (0pt,0pt) rectangle (20pt,20pt);
\end{tikzpicture}}
\hspace{40pt}
\subfloat[]{
\begin{tikzpicture}[baseline={(current bounding box.center)}]
\draw (0pt,0pt) rectangle (20pt,20pt);
\draw[xshift=20pt] (0pt,0pt) rectangle (20pt,20pt);
\draw[yshift=-20pt](0pt,0pt) rectangle (20pt,20pt);
\draw[xshift=40pt] (0pt,0pt) rectangle (20pt,20pt);
\end{tikzpicture}
\hspace{5pt}
\begin{tikzpicture}[baseline={(current bounding box.center)}]
\draw[->,>=stealth](0pt,0pt) -- (40pt,0pt);
\node at (20pt,10pt){reduce};
\end{tikzpicture}
\hspace{5pt}
\begin{tikzpicture}[baseline={(current bounding box.center)}]
\draw (0pt,0pt) rectangle (20pt,20pt);
\draw[xshift=20pt] (0pt,0pt) rectangle (20pt,20pt);
\end{tikzpicture}
}
\caption{(a) Example of reduced Young diagram $\lambda$ in $\operatorname{SU}(N)_2$ and (b) its transpose $\lambda^t$ in $\operatorname{SU}(2)_N$. (We assume $N>3$ here). We can label $\lambda$ by $x=3$, $y=1$ and reduced $\lambda^t$ by $x-y=2$.}
\label{fig: Young diagrams}
\end{figure}
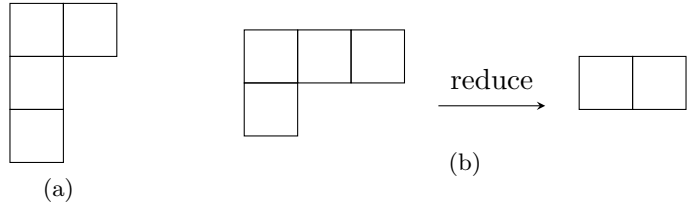

The explicit formula for modular S-matrix of general ${\operatorname{SU}}(N)_k$ is complicated. However, due to the simplicity of $\operatorname{SU}(2)_N$, it is possible to have a simple formula for $\operatorname{SU}(N)_2$, which is given by a level-rank duality on S-matrix (for example, see Ref.~\cite{francesco2012conformal}):
\begin{eqnarray}
s_{\lambda,\mu}=\sqrt{\frac{k}{N}} e^{2\pi i |\lambda| |\mu| /Nk} s^*_{\lambda^t, \mu^t}
\label{eqn: level rank duality for S matrix}
\end{eqnarray}
where reduced Young diagram $\lambda,\mu \in \operatorname{SU}(N)_k$, and their transpose $\lambda^t,\mu^t \in \operatorname{SU}(k)_N$, (e.g., see figure~\ref{fig: Young diagrams} as a demonstration). For $k=2$, diagram $\lambda$ can be conveniently parametrized by two integers: $N-1\geq x \geq y \geq 0$, which count the number of boxes for column 1 and 2. Its transpose is not generally reduced in $\operatorname{SU}(2)_N$, but reduce to a $\lambda^t \in \operatorname{SU}(2)_N $, which can be parametrized by one integer $x-y \in [0, N-1] \cap \mathbb{Z}$. And $|\lambda|$ counts the total number of boxes in reduced diagram: $|\lambda|=x+y$. The identity sector corresponds to $x=y=0$. It is also useful to mention the total number of families in $\operatorname{SU}(N)_k: \frac{(k+N-1)!}{k!(N-1)!}$.

We can derive the modular S-matrix of $\operatorname{SU}(N)_2$ by the explicit formula of the S-matrix of $\operatorname{SU}(2)_N$:
\begin{eqnarray}
s_{(x,y),(x',y')}=\frac{2}{\sqrt{N(N+2)}} e^{2\pi i \frac{(x+y)(x'+y')}{2N}}\sin \left( \frac{(x-y+1)(x'-y'+1)}{N+2} \pi \right)
\end{eqnarray}
Here integers $x,y,x',y'$ satisfy $N-1 \geq x \geq y \geq 0$ and  $N-1 \geq x'\geq y' \geq 0
$. An immediate observation is that when $N\gg 1$, the number of families in $\operatorname{SU}(N)_2$ is much greater than those in $\operatorname{SU}(2)_N$: $N(N+1)/2 \gg N+1$. Therefore, the norm $|s_{\lambda,\mu}| =\sqrt{\frac{2}{N}} |s_{\lambda^t,\mu^t}| $ depends on a smaller set of real numbers $|s_{\lambda^t,\mu^t}|$ in $\operatorname{SU}(2)_N$. 

We can also derive the $|r[\lambda,\mu]|$ of $\operatorname{SU}(N)_2$ from the formula~\ref{eqn: level rank duality for S matrix}:
\begin{eqnarray}
|r[\lambda,\mu]|= \frac{|s_{\lambda^t,\mu^t}| |s_{11}| }{|s_{1,\mu^t}| |s_{1,\lambda^t}|} = |r[\lambda^t,\mu^t]|
\end{eqnarray}
which is identical to those in the dual theory. 

In the following we will show explicitly that not only the spectrum of $|r|$ are identical for the dual theories $\operatorname{SU}(N)_2$ and $\operatorname{SU}(2)_N$, but also the distributions of $|r|$ are identical. $|r[i,j]|$ for two labels $i,j\in \operatorname{SU}(2)_N$, i.e. $i,j=0,1,2,...,N$ has the following expression:
\begin{eqnarray}
|r[i,j]|=\left\lvert  \frac{\sin \left(\frac{\pi}{N+2 }\right)\sin \left(\frac{(i+1)(j+1)\pi}{N+2 }\right)}{\sin \left(\frac{(i+1)\pi}{N+2 }\right)\sin \left(\frac{(j+1)\pi}{N+2 }\right)}  \right\rvert 
\end{eqnarray}

First of all, it can be directly verified that $|r[i,j]|=|r[i,N-j]|$, $\forall i,j$, since
\begin{align}
\left\lvert \sin \left( \frac{(i+1)(j+1)\pi}{N+ 2} \right) \right\rvert 
=  \left\lvert\sin \left[ \left( 1- \frac{(j+1)}{N+ 2} \right) (i+1) \pi \right] \right\rvert 
=  \left\lvert \sin \left(\frac{(N-j+1)(i+1) \pi}{N+ 2} \right) \right\rvert 
\end{align}
Next, we count how many labels in $\operatorname{SU}(N)_2$ are mapped to $j$ and $N-j$ (we assume $j\neq N/2$ for now, and comment later) in $\operatorname{SU}(2)_N$. In general, $j$ in $\operatorname{SU}(2)_N$ corresponds to all pairs of integers $(x,y)$ satisfying $x-y=j,~N-1 \geq x \geq y \geq 0$. There are in total $N-j$ pairs. So together with those corresponds to label $N-j$, there are $N$ channels in $\operatorname{SU}(N)_2$ that correspond to the pair of channels $\lbrace j, N-j \rbrace$ in ${\rm SU}(2)_N$. In other words, for generic two labels $i,j$, there are four identical normed residue values $|r[i,j]|= |r[N-i,j]|= |r[i,N-j]|=|r[N-i,N-j]|$ in $\operatorname{SU}(2)_N$, and there are $N^2$ identical residual values in $\operatorname{SU}(N)_2$. Therefore each $|r[i,j]|$ in $\operatorname{SU}(2)_N$ has $N^2/4$ duplicates in $\operatorname{SU(N)_2}$. For the case when $j=N-j=N/2\in \mathbb{Z}$, there are $N-j=N/2$ labels in $\operatorname{SU}(N)_2$, so that the counting also holds. 

The above counting argument is sufficient to prove that the probability distributions for $|r|$ in $\operatorname{SU}(2)_N$ and $\operatorname{SU}(N)_2$ are identical. We should comment here that we have only discussed the norm of $r$. $r$ has a strongly fluctuating phase, which can distinguish between different $\lambda$'s that have same reduced transpose $\lambda^t \in \operatorname{SU}(2)_N$. This is consistent with the fact that the two dual theories have different average OTOC $\langle r\rangle$ since they have different quantum dimensions. 

In parallel with the observation in $\operatorname{SU}(2)_k$, where scrambling is more sufficient in larger $k$, we can deduce here that scrambling is more sufficient in large $N$ for $\operatorname{SU}(N)_2$ models. Therefore $\operatorname{SU}(N)_2$ is an example of a family of RCFTs with strong scrambling in the large central charge limit, which may be interesting from the point of view of holographic duality. 

\bibstyle{apsrev4-1}
\bibliography{refs}

\end{document}